\documentclass[a4paper,10pt]{article}
\pdfoutput=1 

\usepackage{jheppub} 
\usepackage[T1]{fontenc} 
\usepackage{natbib}
\usepackage{here}
\usepackage{float}
\usepackage{multirow}
\usepackage{graphicx}
\usepackage{subcaption}
\usepackage{multirow}
\usepackage{bbm}
\usepackage{cleveref}
\usepackage[utf8]{inputenc}

\usepackage{hyperref}
\usepackage{bbold}

\usepackage{graphicx,epsf}
\usepackage{amsmath,amsfonts,amssymb,amsbsy}

\usepackage{hyperref}
\usepackage{xcolor}
\hypersetup{
    colorlinks=true,
    allcolors={blue!60!black}
}

\usepackage{multirow}
\usepackage{here}
\usepackage{stackrel}
\usepackage{tikz}
\usetikzlibrary{calc}

\usepackage{ifdraft}
\usepackage[obeyFinal]{easy-todo}

\usepackage{placeins}
\usepackage{slashed} 

\usepackage{booktabs}
\usepackage{adjustbox}

\newcommand{\be}{\begin{equation}}
\newcommand{\ee}{\end{equation}}

\newcommand{\bea}{\begin{eqnarray}}
\newcommand{\eea}{\end{eqnarray}}

\newcommand{\eps}{\epsilon}

\def\fig#1{fig.~{\ref{#1}}}

\def\spa#1.#2{\left\langle#1\,#2\right\rangle}
\def\spb#1.#2{\left[#1\,#2\right]}
\def\spash#1.#2{\spa{\smash{#1}}.{\smash{#2}}}
\def\spbsh#1.#2{\spb{\smash{#1}}.{\smash{#2}}}
\def\sand#1.#2.#3{%
\left\langle\smash{#1}{\vphantom1}^{-}\right|{#2}%
\left|\smash{#3}{\vphantom1}^{-}\right\rangle}
\def\sandpp#1.#2.#3{%
\left\langle\smash{#1}{\vphantom1}^{+}\right|{#2}%
\left|\smash{#3}{\vphantom1}^{+}\right\rangle}
\def\sandpm#1.#2.#3{%
\left\langle\smash{#1}{\vphantom1}^{+}\right|{#2}%
\left|\smash{#3}{\vphantom1}^{-}\right\rangle}
\def\sandmp#1.#2.#3{%
\left\langle\smash{#1}{\vphantom1}^{-}\right|{#2}%
\left|\smash{#3}{\vphantom1}^{+}\right\rangle}

\def\trFive{{\rm tr}_5}

\def\offShellScale{p_1^2}
\def\offShellScaleSquared{p_1^4}

\def\dlog{$\mathrm{d}\log$}

\def\trFive{{\rm tr}_5}

\newcommand*{\zmz}{\textrm{zmz}}
\newcommand*{\mzz}{\textrm{mzz}}
\newcommand*{\zzz}{\textrm{zzz}}

\newcommand*{\nsqrt}{\Sigma_5}

\title{
    Two-Loop Hexa-Box Integrals for Non-Planar Five-Point One-Mass Processes
}

\preprint{CERN-TH-2021-114, FR-PHENO-2021-009}

\author[1,2,3]{Samuel Abreu,}

\affiliation[1]{Theoretical Physics Department, CERN, Geneva, Switzerland}
\affiliation[2]{Mani L. Bhaumik Institute for Theoretical Physics,  Department of Physics and Astronomy,\\
UCLA, Los Angeles, CA 90095, USA}
\affiliation[3]{Higgs Centre for Theoretical Physics, School of Physics and Astronomy,\\
The University of Edinburgh, Edinburgh EH9 3FD, Scotland, UK}

\author[4]{Harald Ita,}
\affiliation[4]{Physikalisches Institut, Albert-Ludwigs-Universit\"at Freiburg, \\
D-79104 Freiburg, Germany}

\author[1]{Ben Page,}

\author[4]{Wladimir Tschernow}

\abstract{
We present the calculation of the three distinct non-planar
hexa-box topologies for five-point one-mass processes.
These three topologies are required to obtain the two-loop virtual QCD
corrections for two-jet-associated W, Z or Higgs-boson production.
Each topology is solved by obtaining a pure basis of master integrals
and efficiently constructing the associated differential equation with numerical
sampling and unitarity-cut techniques. We present
compact expressions for the alphabet of these non-planar integrals,
and discuss some properties of their symbol. Notably, we observe
that the extended Steinmann relations are in general not satisfied.
Finally, we solve the differential equations in terms of generalized power
series and provide high-precision
values in different regions of phase space which can be used as boundary
conditions for subsequent evaluations.
}

\begin{document}

\maketitle


\section{Introduction}

The evaluation of Feynman integrals is a central problem one needs
to address when computing loop corrections in any perturbative quantum
field theory (QFT). Modern approaches to the calculation of loop corrections
to various quantities in QFT proceed by decomposing them into
a sum of products of algebraic coefficients and master integrals.
The algebraic coefficients depend on the loop order, the quantity being computed
and the QFT under consideration.
The master integrals, on the other hand, only depend on the 
underlying kinematics and loop order. The evaluation of master integrals
is thus an interesting problem on its own.

Despite receiving a lot of attention in recent years, the calculation
of Feynman integrals still poses a major challenge in obtaining
two-loop corrections for processes of great physical interest. 
At the multi-leg and multi-loop frontier, 
the complete set of master integrals required
for the scattering of five massless particles has been 
computed~\cite{Gehrmann:2015bfy,Papadopoulos:2015jft,Abreu:2018aqd,
Chicherin:2018old,Chicherin:2020oor}, which has led to a large
number of new analytic two-loop amplitudes in both supersymmetric theories
and in QCD~\cite{Abreu:2018aqd,Chicherin:2018yne,Abreu:2019rpt,Chicherin:2019xeg,Badger:2018enw,Abreu:2018zmy,Abreu:2019odu,Abreu:2021oya,Badger:2019djh,Chawdhry:2019bji,Abreu:2020cwb,Chawdhry:2020for,Agarwal:2021grm,Chawdhry:2021mkw,Badger:2021imn}.

More recently, all planar master integrals relevant for the scattering 
of a massive particle and four-massless ones were
computed~\cite{Abreu:2020jxa,Canko:2020ylt}. These results have already allowed
the calculation of the two-loop QCD corrections to the 
$u\bar{d}\to W^+b\bar{b}$ process~\cite{Badger:2021nhg} at leading color,
as well as the calculation of new two-loop form factors in 
planar $\mathcal{N}=4$ SYM~\cite{Guo:2021bym}. However,
for other important physical processes at hadron colliders, such as
the production of a Z or Higgs boson in association with two jets, 
as well as sub-leading color effects in the production of a W boson in
association with two jets,
the planar integrals considered in refs.~\cite{Abreu:2020jxa,Canko:2020ylt}
are not sufficient.
To extend the results of ref.~\cite{Abreu:2020jxa} beyond the planar
limit there are five new topologies of master integrals that must be computed.
These can be grouped into two sets: there are three distinct
hexa-box topologies, and two distinct double-pentagon topologies.
While one of the hexa-box topologies has been considered
previously~\cite{Papadopoulos:2019iam}, 
in this paper we compute for the first time the full set of 
hexa-box topologies.

Our calculation of the hexa-box master integrals follows
the approach used in ref.~\cite{Abreu:2020jxa},
which gives us detailed insight into the analytic
structure of the master integrals but also allows one to 
numerically evaluate the master integrals at any phase-space
point.
We start by constructing differential equations for the master 
integrals~\cite{Kotikov:1990kg,Kotikov:1991pm, Bern:1993kr, 
Remiddi:1997ny, Gehrmann:1999as}, choosing a basis of `pure'
master integrals~\cite{ArkaniHamed:2010gh} so that
the differential equations take a particularly simple `canonical' 
form~\cite{Henn:2013pwa}, which only involves \dlog{} forms.
In practice, we construct the basis with a heuristic approach that is validated
by constructing the differential equation and observing the canonical form.
Contrary to the massless five-point case~\cite{Chicherin:2017dob}, we find that
the collection of \dlog{} forms which appear in the differential equations is not
given by the permutation closure of those arising in the planar
topologies~\cite{Abreu:2020jxa}.
Indeed, we find a new class of square root, not associated to a momentum-space gram
determinant.
In order to determine the remaining \dlog{} forms, we use the approach of
ref.~\cite{Abreu:2020jxa} and consider much simpler differential equations
where the propagators are put on shell.
To obtain the analytic form of the differential equation
we use the numerical sampling method of ref.~\cite{Abreu:2018rcw}, which
trivializes the integral reduction of the differential equations
and can be implemented over finite fields~\cite{vonManteuffel:2014ixa, 
Peraro:2016wsq}.

Once the canonical differential equation is known, it is trivial
to determine the so-called `symbol'~\cite{Goncharov:2010jf,Duhr:2011zq,
Duhr:2012fh} of the master integrals, which gives valuable
insight into the analytic structure of the master integrals
in a compact format. 
In a nutshell, these integrals evaluate to multi-valued functions with
complicated branch-cut structures, and the symbol encodes the information 
about the position of all logarithmic singularities. The symbol
is built out of `letters', which are algebraic functions of the kinematic
variables that vanish at the logarithmic singularities of the integrals.
The symbol contains non-trivial information about the analytic
properties of the master integrals and scattering amplitudes.
For instance, the analytic form
of the letters can be used to greatly simplify the calculation of
loop amplitudes~\cite{Abreu:2018zmy,Abreu:2019odu}.
The symbols also encode the discontinuities of Feynman integrals,
which are constrained by physical considerations.
As expected, we observe that the symbols of the hexa-box integrals
satisfy the Steinmann 
relations~\cite{Steinmann, Steinmann2,Cahill:1973qp, Caron-Huot:2016owq,
Dixon:2016nkn}.  However, interestingly, we find that the 
`extended Steinmann relations'~\cite{Caron-Huot:2018dsv} are
in general not satisfied.

The differential equation can also be used to numerically evaluate the master
integrals. One approach would be to analytically solve the differential
equation in terms of multiple polylogarithms (MPLs). However, given the high number of variables and the weight of the MPLs, for the hexa-box integrals
we consider this would lead to lengthy expressions which are
specific to a given phase-space region and highly non-trivial to analytically continue.
Instead, we numerically solve
the differential equations using generalized power 
series~\cite{Francesco:2019yqt,Hidding:2020ytt}, a method which was already
shown to be suitable for the planar two-loop five-point one-mass 
integrals~\cite{Abreu:2020jxa}. Compared to Monte-Carlo based approaches
\cite{Smirnov:2015mct,Borowka:2017idc,
Capatti:2019ypt,Capatti:2019edf,Runkel:2019yrs}, the numerical solution
of the differential equation in terms of generalized power series
allows to obtain high-precision results 
(see also the approaches of refs.~\cite{Mandal:2018cdj,Liu:2021wks}), 
which, for instance, means that we can evaluate the integrals in 
singular regions of phase space to arbitrary precision.
To obtain numerical values for the hexa-box integrals, we determine
a set of initial conditions by requiring that the integrals are free of spurious
branch cuts in their Euclidean region. These initial
conditions correspond to the value of the integrals at a Euclidean
phase-space point, and they were obtained to more than 100 digits
with an in-house code. We then use the code of ref.~\cite{Hidding:2020ytt}
to obtain high-precision numerical values in different regions of
phase space, which can themselves be used as initial conditions
for subsequent evaluations.

To facilitate the use of our results, we include a set of ancillary
files which contain all our analytic and numerical results.
For each of the three hexa-box topologies, we include a file
with the definition of the pure basis, \texttt{anc/*/*\_pure\_basis.m},
and a file with a graphical representation of the master integrals
\texttt{anc/*/*\_graphs.m}. The differential equations can be assembled
from the list of matrices in \texttt{anc/*/*\_connection.m} and the
alphabet in \texttt{anc/alphabet.m}. We also include the
file \texttt{anc/usageExample.m} which illustrates the use of the ancillary
files, and allows to compute the symbols of all master integrals to
the desired weight.

The paper is structured as follows. In section \ref{sec:scatKin}
we describe the kinematics relevant for the master integrals we will
be computing and introduce quantities such as Gram determinants that will be
important for the construction of the differential equations.
In section \ref{sec:masterInt} we define the hexa-box topologies.
Next, in section \ref{sec:pureBasisConstruction} we discuss the canonical differential
equations and present the pure bases for each topology.
In section \ref{sec:DifferentialEquations} we discuss the symbol alphabet,
the analytic form of the differential equations and some properties of the
symbols of the hexa-box integrals.
Finally, in section \ref{sec:numerics} we discuss the numerical 
solution of the differential equations, before we present our 
conclusions in section \ref{sec:Conclusions}.


\section{Scattering Kinematics and Notation}
\label{sec:scatKin}

We consider the scattering of five particles, four of which are massless.
We denote their momenta by $p_i,$ $i=1,\ldots,5$, which
satisfy the momentum conservation $\sum_{i=1}^5 p_i =0$ and, without loss of
generality, we take $p_1^2 \neq 0$, and $p_{i}^2=0$ for  $i=2,\ldots,5$.
The Mandelstam variables $s_{ij}=(p_i+p_j)^2$ for arbitrary $i$ and $j$
can all be written as linear combinations of the six variables
\begin{equation} \label{eqn:orderedInvariants}
\vec s= \{ \offShellScale{} \,,
s_{12}\,, s_{23}\,,s_{34}\,, s_{45}\,, s_{15}\}\,.
\end{equation}
These six variables are however not sufficient to specify a
point in the five-particle phase space. Indeed, this space is separated into
two halves which are mapped onto each other by a space-time
parity transformation. The parity label of each point can be captured
by the parity-odd contraction with the Levi-Civita tensor
\begin{equation}
\label{eq:tr5}
\trFive{} = 4 i \varepsilon_{\alpha\beta\gamma\delta} 
\,p_1^\alpha p_2^\beta p_3^\gamma p_4^\delta\,.
\end{equation}

A ubiquitous set of quantities that appears when describing the kinematics
of a scattering process are the Gram determinants one can form with 
(subsets of) the momenta $p_i$.
They are given by the determinants of the
Gram matrix $G(q_1,\ldots,q_n)$, which we define as
\begin{equation}\label{eq:gramGen}
	G(q_1,\ldots,q_n)=2\,V^T(q_1,\ldots,q_n)\,g\,V(q_1,\ldots,q_n)
	=2 \, \{q_i\cdot q_j\}_{i,j\in\{1,\ldots,n\}}\,,
\end{equation}
where  $V(q_1,\ldots,q_{n})$ is 
a $4\times n$ matrix whose columns are the vectors $q_i$.
For concreteness, in this paper we use the metric $g = \mathrm{diag}(+,-,-,-)$,
which we extend with further minus signs when working with $D$-dimensional
momenta.
It is clear from \cref{eq:gramGen} that the Gram determinants are
just polynomials in the Mandelstam variables. 
We will be particularly interested in the three-point Gram determinant
\begin{align}\begin{split}
	\label{eq:gram3}
	\Delta_3=&\,-\det G(p_1, p_2+p_3)\\
	=&\,s_{23}^2+s_{45}^2+\offShellScaleSquared-2 s_{23} s_{45}-2 \offShellScale{} s_{23}-2 \offShellScale{} s_{45}\\
	=&\,\lambda(\offShellScale{}, s_{23},s_{45})\,,
\end{split}\end{align}
where $\lambda(a,b,c)=a^2+b^2+c^2-2ab-2ac-2bc$ is the
K\"all\'en function, and in the five-point Gram determinant
\begin{align}\begin{split}
	\label{eq:gram5}
	\Delta_5 =&\,  \det G(p_1,p_2,p_3,p_4)\\
	=&\,(s_{12} s_{15} - s_{12} s_{23} - \offShellScale{} s_{34} 
	- s_{15} s_{45} + s_{34} s_{45} + s_{23} s_{34})^2	\\
	&\,- 4 s_{23}  s_{34} s_{45} (\offShellScale{}-s_{12} - s_{15} + s_{34})\,.
\end{split}\end{align}
The five-point Gram determinant $\Delta_5$ is closely related
to the $\trFive{}$ defined in \cref{eq:tr5}, as we have that
\begin{equation}
	\Delta_5=\trFive^2\,.
\end{equation}
It is then very tempting to identify $\trFive{}$ with $\sqrt{\Delta_5}$, but
one must take care with this identification.
While $\trFive{}$ is a square root of $\Delta_5$, the branch
choice (correspondingly the sign) is fixed by the underlying set of momenta defining the
phase-space point. In contrast, the branch choice/sign of $\sqrt{\Delta_5}$ can
conveniently be fixed by the standard prescription of the square-root map.
With this convention $\sqrt{\Delta_5}$ depends only on the $s_{ij}$ themselves and not on $\trFive{}$.
Consequently, $\trFive{}$ is odd under parity transformations while
 $\sqrt{\Delta_5}$
is invariant, and to relate $\trFive{}$ to $\sqrt{\Delta_5}$ we must encode the
different choices of branch of the square root.
To avoid these complications, and because it is sufficient for the
purpose of this paper, we will avoid referring to $\trFive{}$ and instead
use $\Delta_5$ and $\sqrt{\Delta_5}$.

Another quantity that will appear throughout this paper and
which is closely related to $\Delta_5$ is
\begin{align}\begin{split}\label{eq:newRoot}
	\nsqrt=&\,(s_{12} s_{15} - s_{12} s_{23}  
	- s_{15} s_{45} + s_{34} s_{45} + s_{23} s_{34})^2	
	- 4 s_{23}  s_{34} s_{45} (s_{34} - s_{12} - s_{15})\,.
\end{split}\end{align}
In particular we note that in the limit where $p_1^2\to0$ the two quantities
coincide, that is
\begin{equation*}
\nsqrt=\Delta_5 \qquad \text{for}\qquad p_1^2=0\,,
\end{equation*}
as is manifest from \cref{eq:gram5,eq:newRoot}. We will see below
how this polynomial appears in the construction of the pure basis.

Given a phase-space point $P=\{p_1,p_2,p_3,p_4,p_5\}$, it is natural
to consider the points obtained by permuting the massless legs.
More formally, let us consider the different points 
$\sigma (P)=\{p_{1},p_{\sigma(2)},p_{\sigma(3)},p_{\sigma(4)},p_{\sigma(5)}\}$
where $\sigma\in S_4$ corresponds to a permutation of $\{2,3,4,5\}$.
The action of $\sigma$ on the $s_{ij}$ is trivially defined as
$\sigma (s_{ij})=(p_{\sigma(i)}+p_{\sigma(j)})^2$.
While $\Delta_5$ is invariant
under these permutations,\footnote{As is $\sqrt{\Delta_5}$, 
but we note that $\trFive{}$ is not!} $\Delta_3$ is not. We find that there
are three independent permutations of $\Delta_3$, namely
\begin{equation}
	\Delta_3^{(1)}=\lambda(\offShellScale{}, s_{23},s_{45})\,,\qquad
	\Delta_3^{(2)}=\lambda(\offShellScale{}, s_{24},s_{35})   \qquad\text{and}\qquad
	\Delta_3^{(3)}=\lambda(\offShellScale{}, s_{25},s_{34})\,.
\end{equation}
The polynomial $\nsqrt$ defined in \cref{eq:newRoot} is also
not invariant under a general $\sigma\in S_4$, and we find that it
appears in six different permutations, which we denote
by $\nsqrt^{(k)}$, for $k=1,\ldots,6$, with $\nsqrt^{(1)}=\nsqrt$.


\section{Hexa-box Topologies}
\label{sec:masterInt}
\begin{figure}
	\centering
	\begin{subfigure}{0.3\textwidth}\centering
		\includegraphics[scale=0.3]{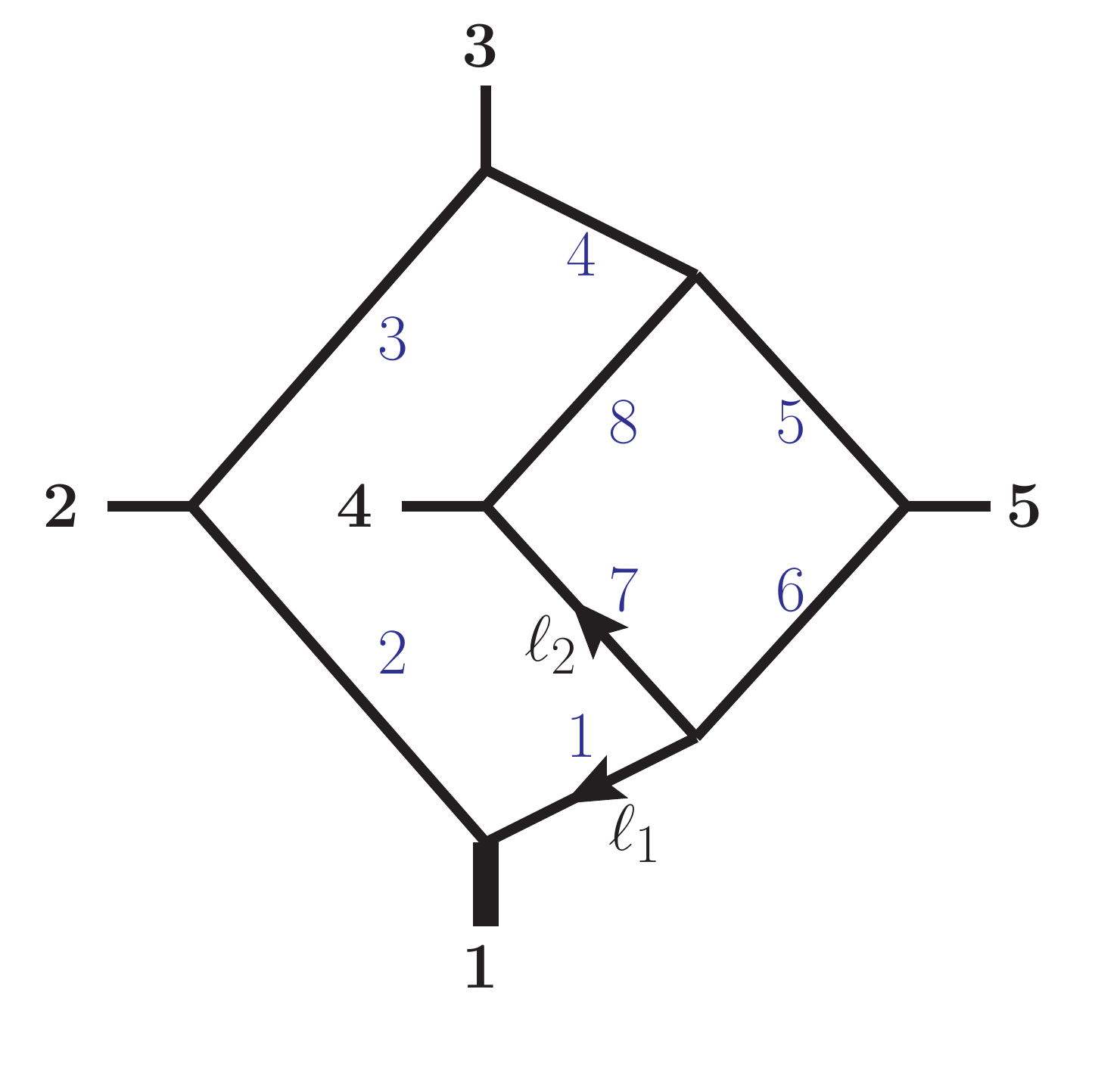}
		\caption{$I_{\rm hb}^{[\mzz]}[\vec\nu]$}
		\label{fig:mzz}
	\end{subfigure}
	\begin{subfigure}{0.3\textwidth}\centering
		\includegraphics[scale=0.3]{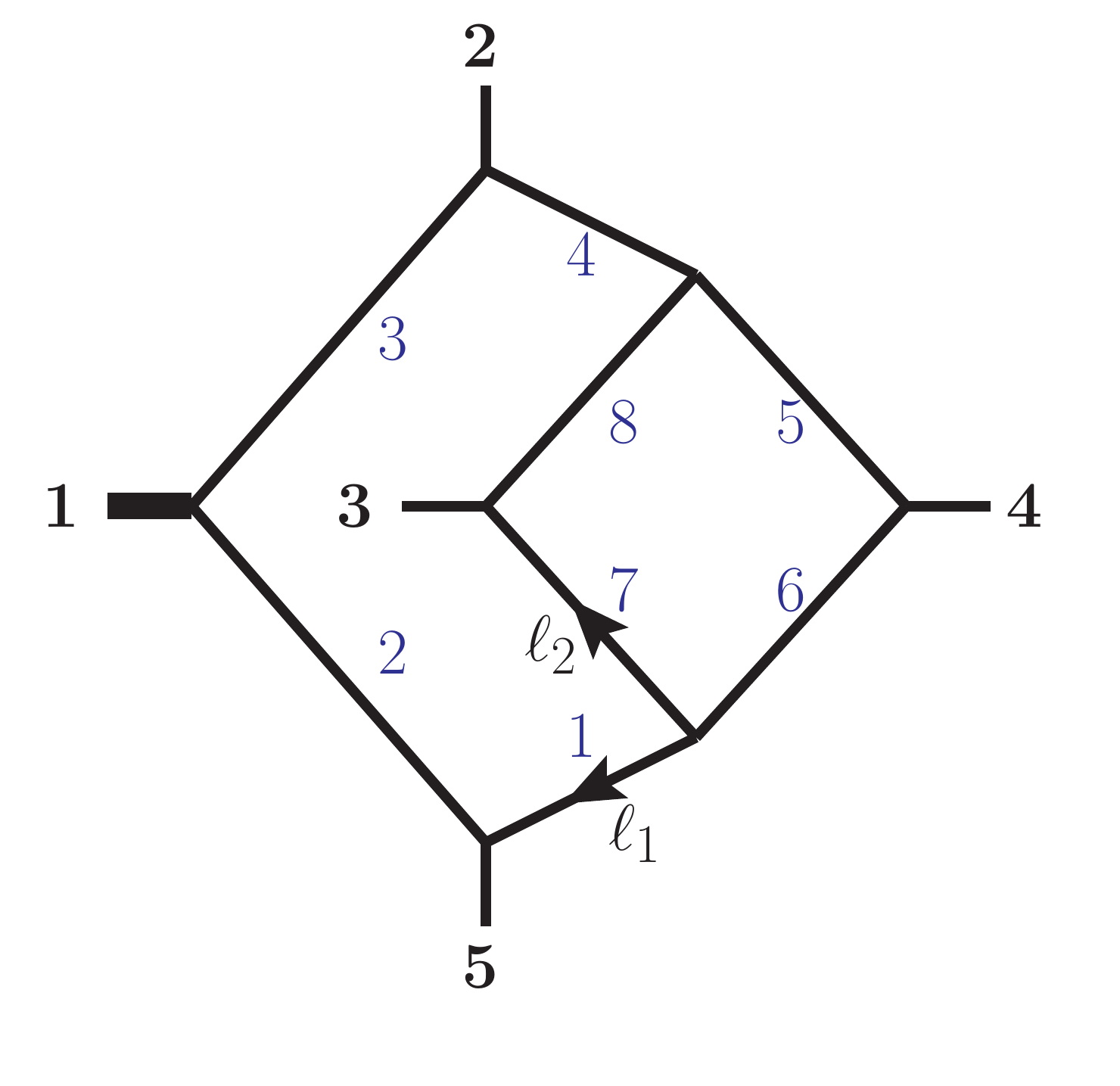}
		\caption{$I_{\rm hb}^{[\zmz]}[\vec\nu]$}
		\label{fig:zmz}
	\end{subfigure}
	\begin{subfigure}{0.3\textwidth}\centering
		\includegraphics[scale=0.3]{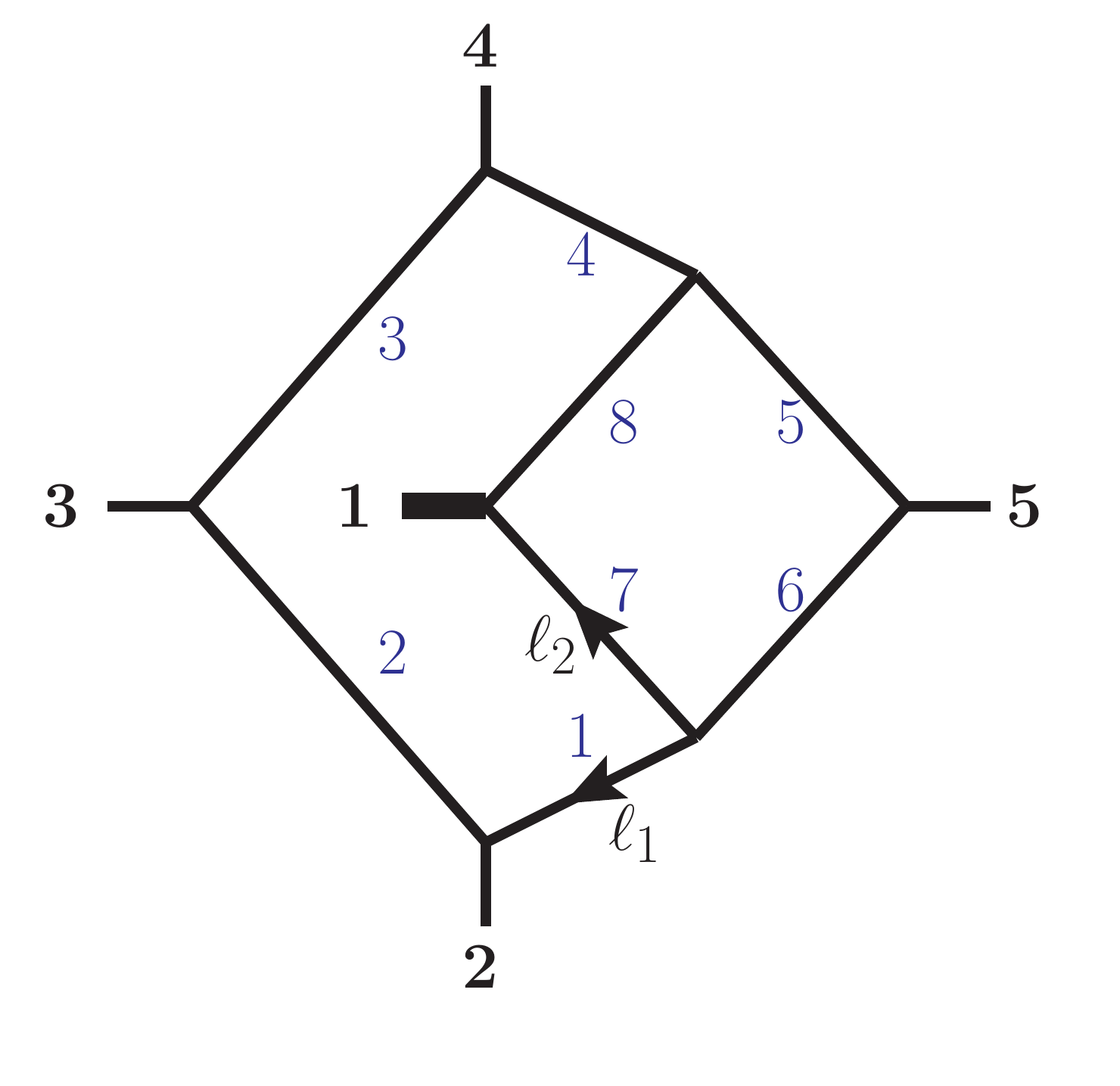}
		\caption{$I_{\rm hb}^{[\zzz]}[\vec\nu]$}
		\label{fig:zzz}
	\end{subfigure}
	\caption{Two-loop five-point one-mass non-planar hexa-box topologies. The thick external line
	with label 1 denotes the massive external leg.} 
	\label{fig_families_int}
\end{figure}

There are three non-planar hexa-box topologies with a single
massive external leg that
are not related by a relabelling of the external momenta.
These three topologies are depicted in \fig{fig_families_int}.
We denote them by $I_{\rm hb}^{[f]}$, with $f\in\{\mzz,\zmz,\zzz\}$ 
distinguishing the mass assignment for the three external legs of the 
hexagon sub-loop: 
they can all have zero mass ($\zzz$), the middle leg can be 
massive ($\zmz$), or the first leg can be massive ($\mzz$).
Each topology defines a linear space $Y^{[f]}$ of integrals 
corresponding to linear combinations of integrals of the form
\begin{equation}\label{eq:intHexaboxes}
I_{\rm hb}^{[f]}[\vec \nu] =  
\int \frac{{\rm d}^D\ell_1}{i\pi^{D/2}}\frac{{\rm d}^D\ell_2 }{i\pi^{D/2}} 
\frac{\rho_{9,f}^{-\nu_9}\,\,\rho_{10,f}^{-\nu_{10}}\,\,\rho_{11,f}^{-\nu_{11}}}
{\rho_{1,f}^{\nu_1}\,\,\rho_{2,f}^{\nu_2}\,\,
\rho_{3,f}^{\nu_3}\,\,\rho_{4,f}^{\nu_4}\,\,
\rho_{5,f}^{\nu_5}\,\,\rho_{6,f}^{\nu_6}\,\,
\rho_{7,f}^{\nu_7}\,\,\rho_{8,f}^{\nu_8}}\,,
\end{equation}
where we set $D=4-2\epsilon$.
Each element in this spanning set is distinguished
by a set of integer values $\vec \nu$, with the restriction
that $\nu_9,\nu_{10},\nu_{11} \leq 0$.
Concretely, the propagator variables ($\rho_{i,f},\,i=1,\ldots, 8$) and the irreducible scalar products  
($\rho_{i,f},\,i = 9,\dots,11$) are 
defined for each topology as 
\begin{equation}\label{eq:propvars}
   \begin{split}
    \vec \rho_\mzz=&\left\{\ell_1^2,(\ell_1+p_1)^2,(\ell_1+p_1+p_2)^2,(\ell_1+p_1+p_2+p_3)^2,(\ell_1+\ell_2-p_5)^2,(\ell_1+\ell_2)^2,\right. \\
  & \quad \left. \ell_2^2,(\ell_2+p_4)^2, (\ell_2+p_1)^2,(\ell_1+p_4)^2,(\ell_2+p_1+p_2)^2 \right\},\\[2mm]
  \vec \rho_\zmz=&\left\{\ell_1^2,(\ell_1+p_5)^2,(\ell_1+p_5+p_1)^2,(\ell_1+p_5+p_1+p_2)^2,(\ell_1+\ell_2-p_4)^2,(\ell_1+\ell_2)^2,\right. \\
  &  \quad \left. \ell_2^2,(\ell_2+p_3)^2, (\ell_2+p_5)^2,(\ell_1+p_3)^2,(\ell_2+p_5+p_1)^2 \right\},\\[2mm]
     \vec \rho_\zzz=&\left\{\ell_1^2,(\ell_1+p_2)^2,(\ell_1+p_2+p_3)^2, (\ell_1+p_2+p_3+p_4)^2,(\ell_1+\ell_2-p_5)^2,(\ell_1+\ell_2)^2,\right. \\
  & \quad  \left.\ell_2^2,(\ell_2+p_1)^2, (\ell_2+p_2)^2,(\ell_1+p_1)^2,(\ell_2+p_2+p_3)^2 \right\}.
    \end{split}
\end{equation}
In \fig{fig_families_int}, where we assume that all external momenta are 
incoming, we include the index associated with
each propagator and the routing of the loop momenta $\ell_1$ and $\ell_2$.

The integrals specified in \cref{eq:intHexaboxes} define a space of integrals
$Y^{[f]}$ associated with each topology. In this paper
we compute a basis of these spaces, i.e., a set of master integrals associated
with each topology. The projection of any element of this space onto the basis
of master integrals can be algorithmically constructed with integration-by-parts
(IBP)
identities~\cite{Chetyrkin:1981qh,Tkachov:1981wb,Laporta:2001dd}.
The dimensions of the bases are
\begin{equation}
 \begin{split}\label{eq:masterCount}
 &\text{dim}(Y^{[\mzz]})=86,\ \ \ \  \text{dim}(Y^{[\zmz]})=86,\ \ \ \  \text{dim}(Y^{[\zzz]})=135.
 \end{split}
\end{equation}
While it is trivial to find some basis for each of these
spaces, one of the main results of this paper will
be the construction of pure bases, which have particularly
nice properties. This will be discussed in detail in the next
section.

\begin{figure} \centering\begin{tikzpicture}[scale=1.1]
	  \node at (2.8,1){\includegraphics[scale=0.15]{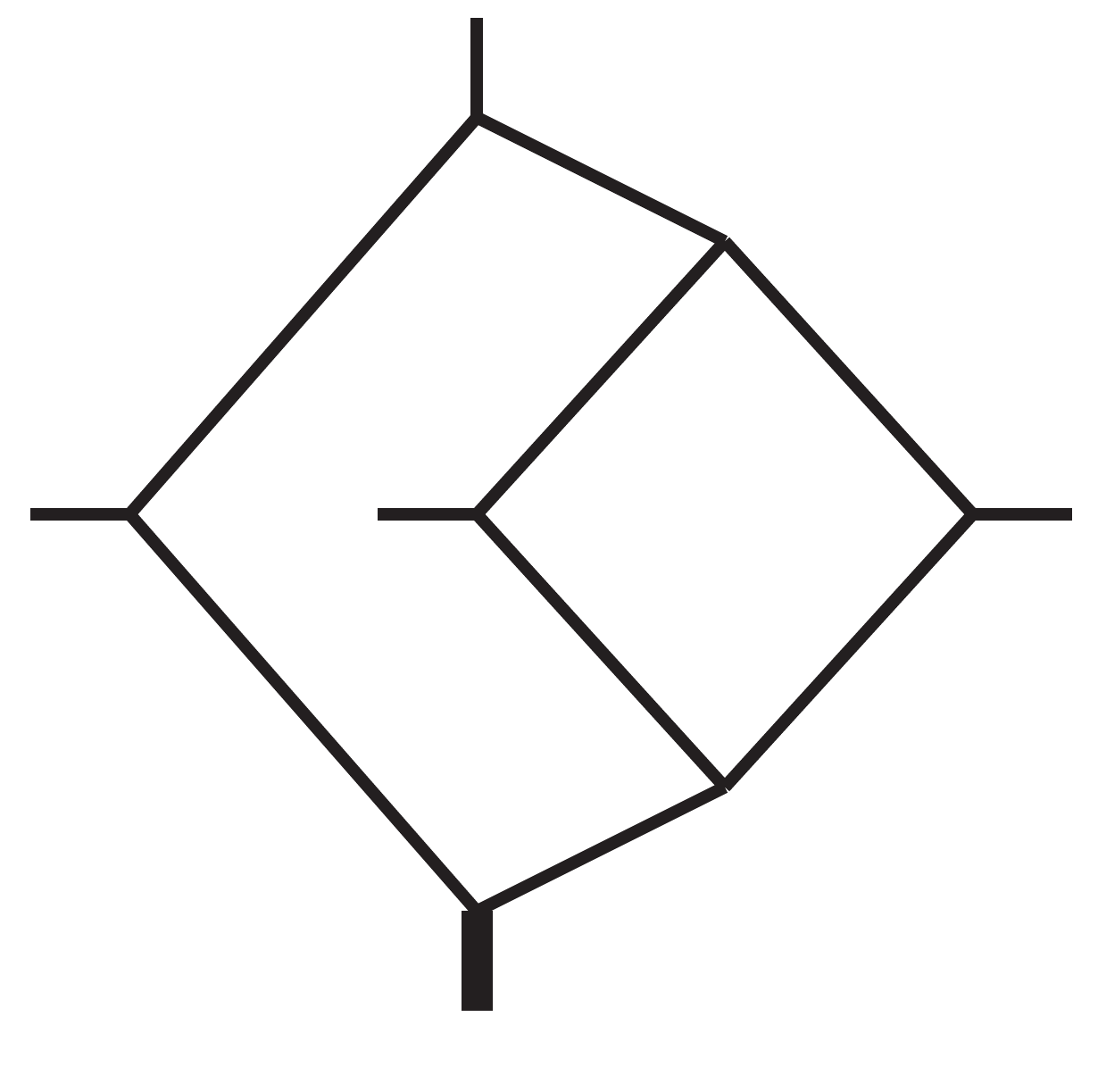}};
		\node at (2.8,0){3 masters}; \node at
		(6.3,1){\includegraphics[scale=0.15]{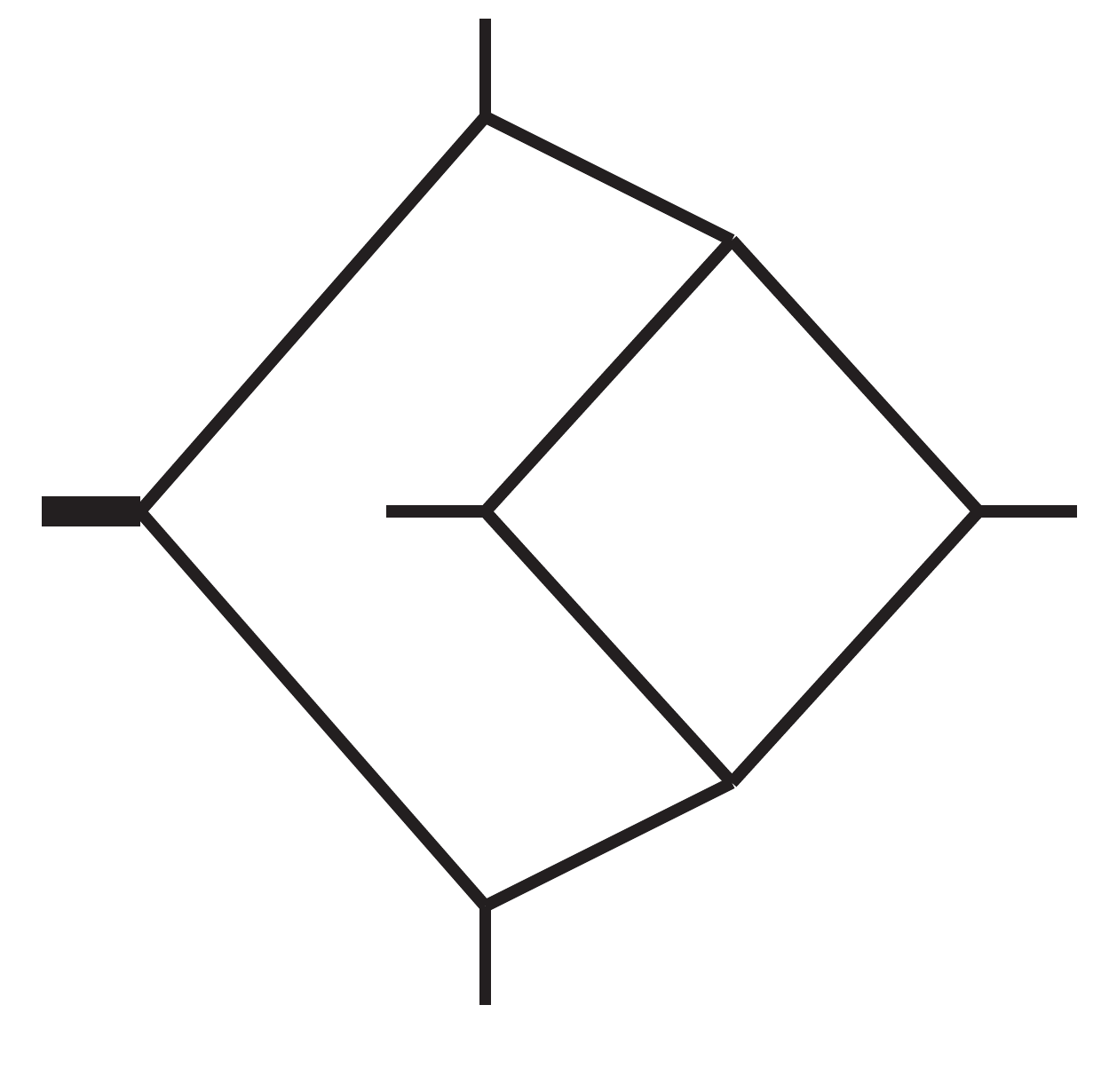}}; \node at
		(6.3,0){3 masters}; \node at
		(9.8,1){\includegraphics[scale=0.15]{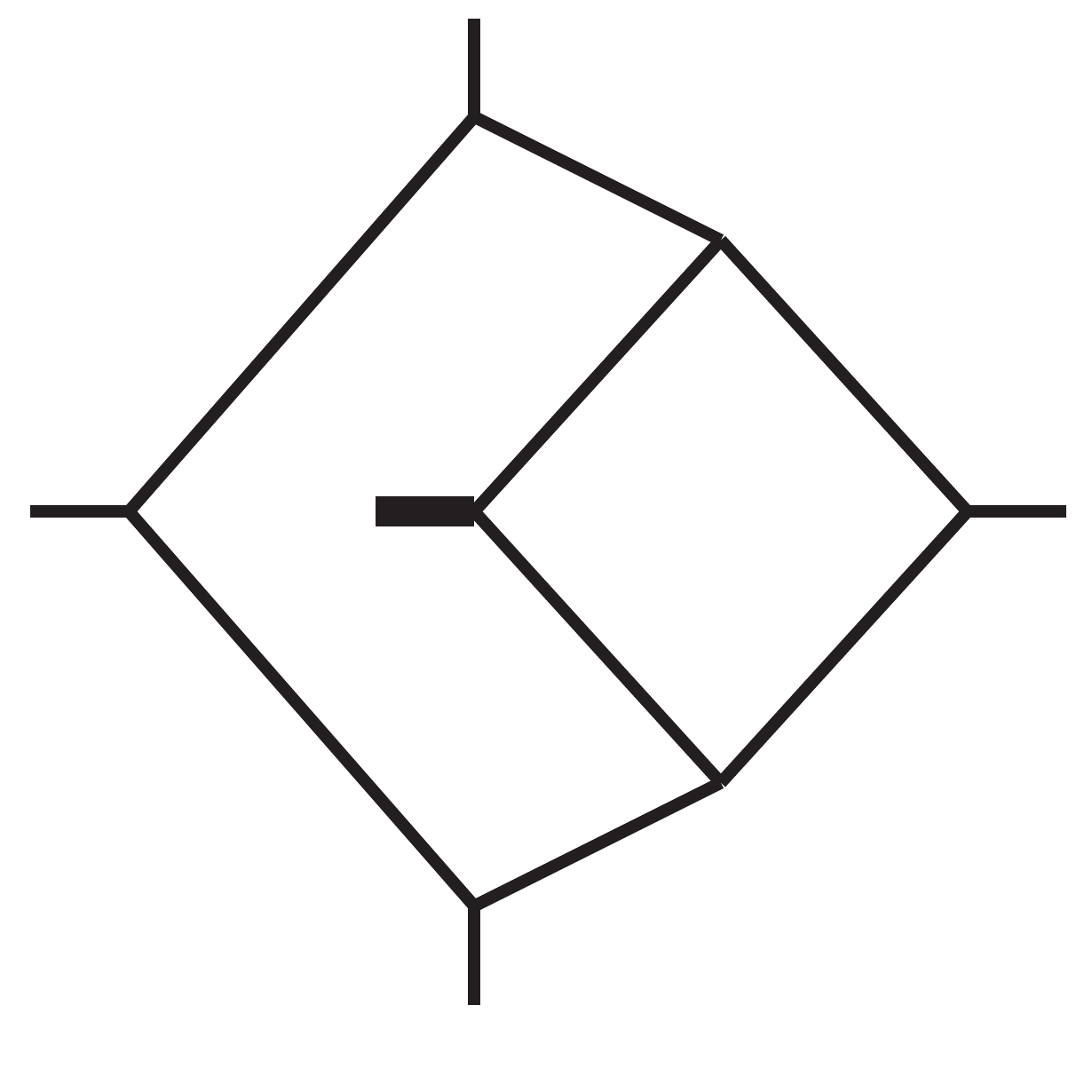}}; \node
		at (9.8,0){3 masters};
	\node at (1.5,-1.5){\includegraphics[scale=0.15]{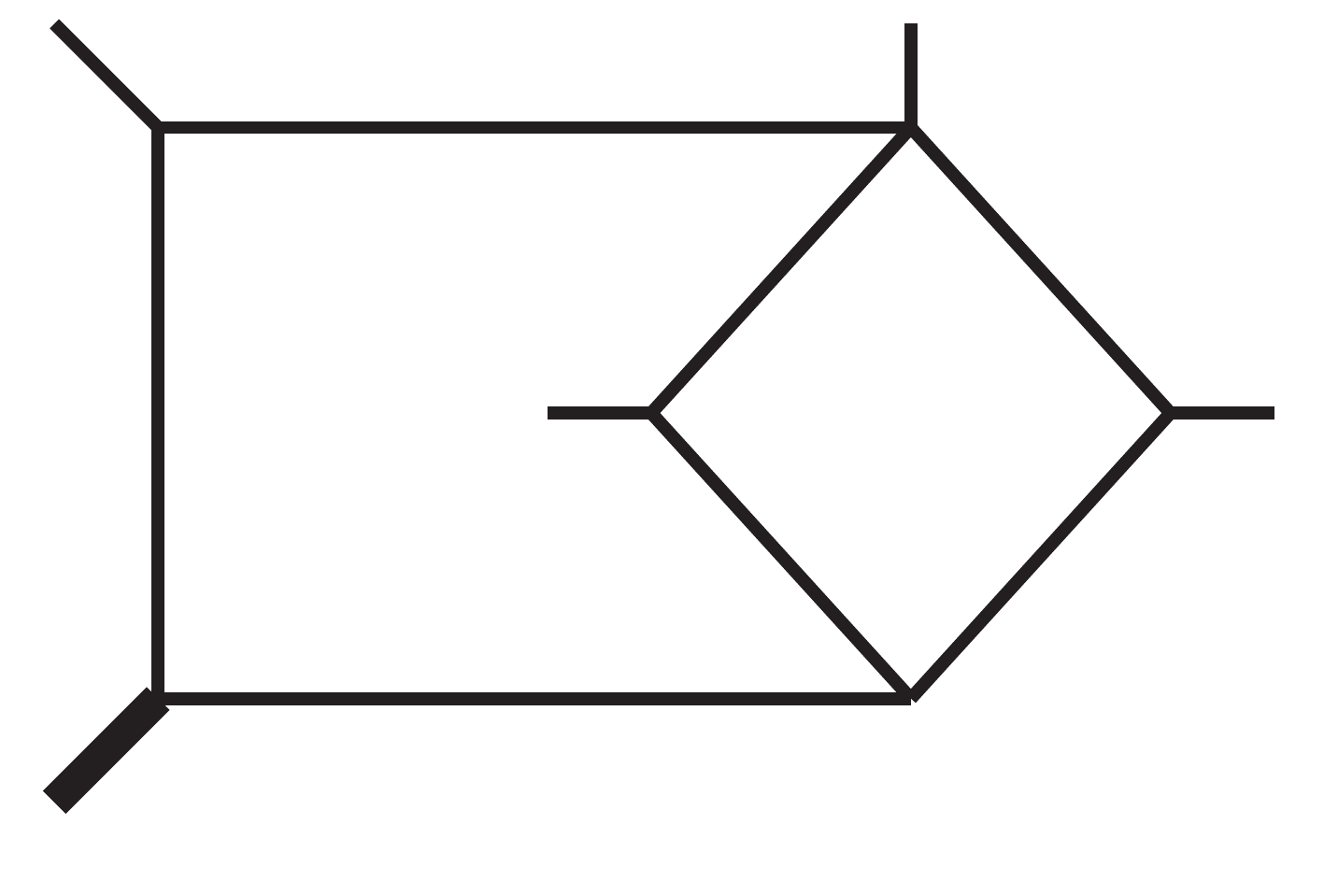}};
	\node at (1.5,-2.3){3 masters}; 
	\node at (5,-1.5){\includegraphics[scale=0.15]{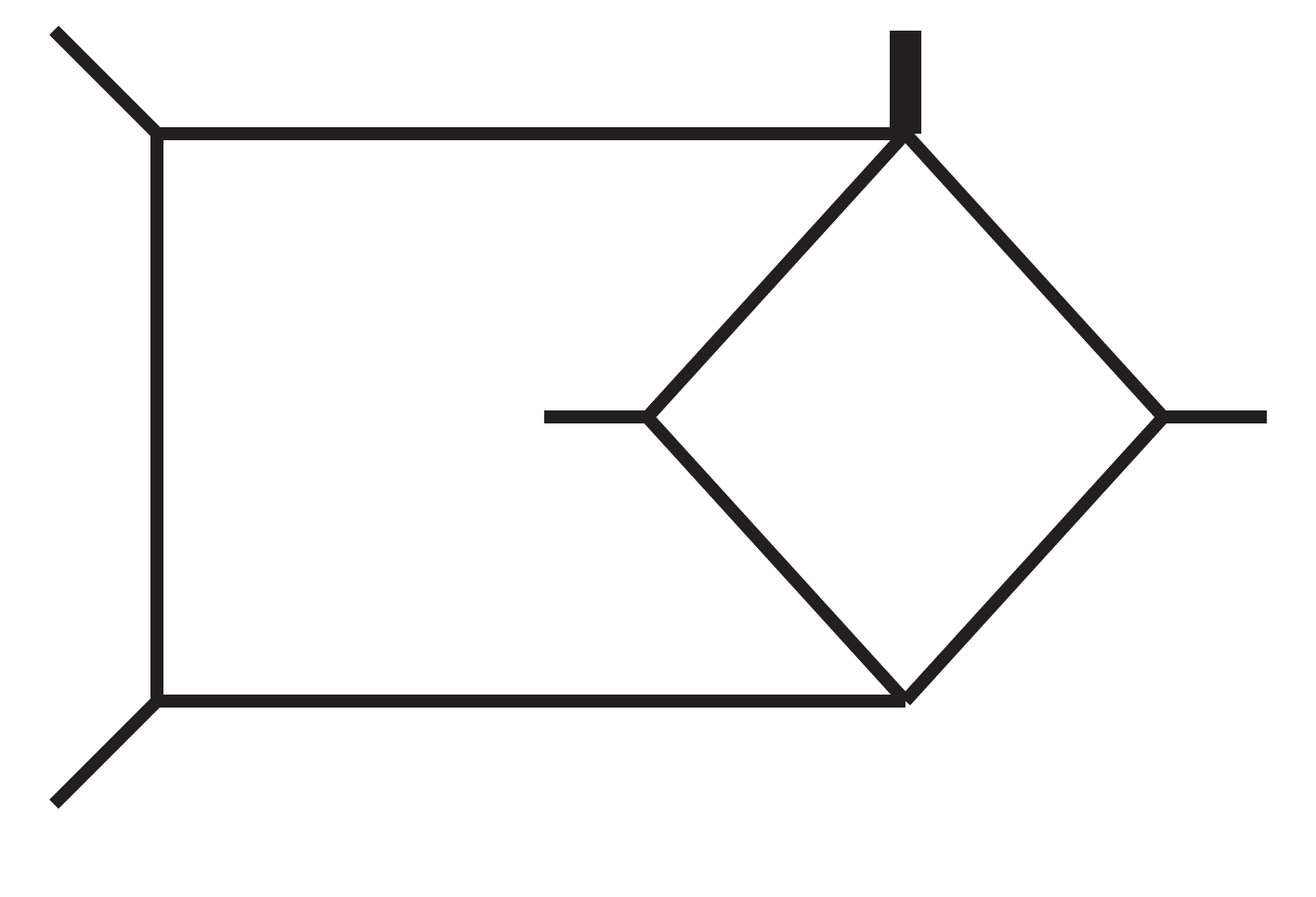}}; 
	\node at (5,-2.3){3 masters}; 
	\node at (8.5,-1.5){\includegraphics[scale=0.15]{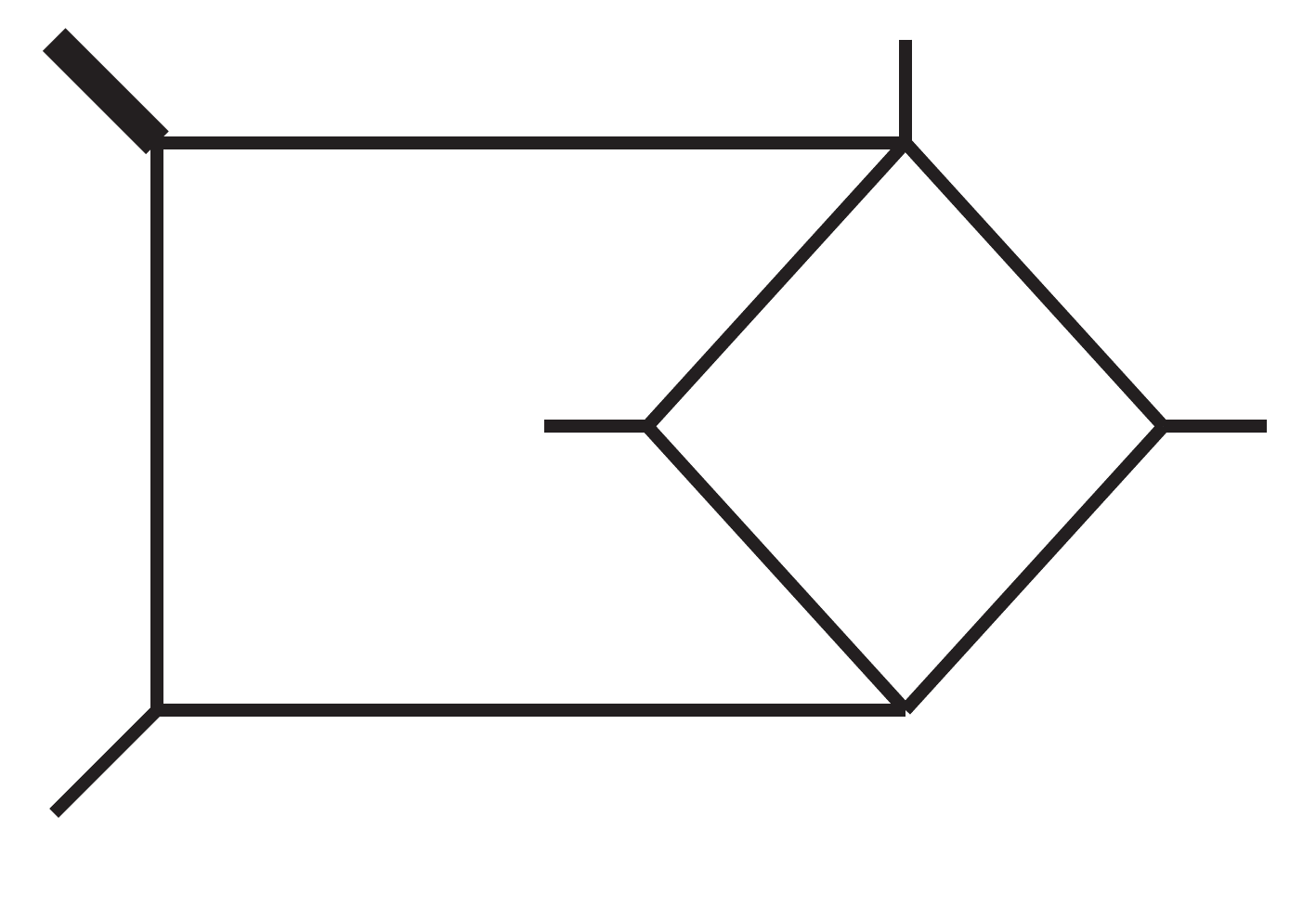}}; 
	\node at (8.5,-2.3){3 masters};
    \node at (12,-1.5){\includegraphics[scale=0.15]{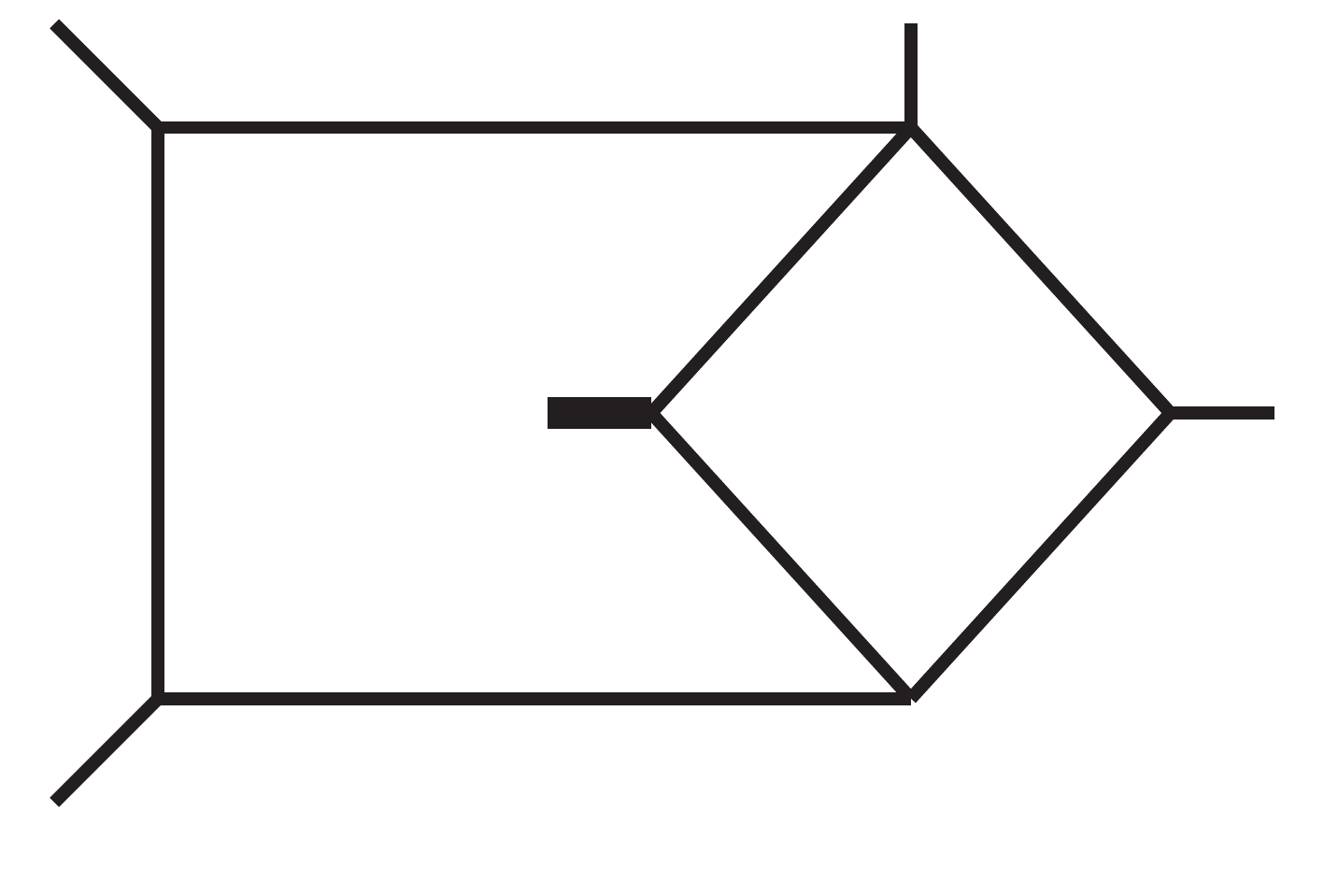}}; 
    \node at (12,-2.3){6 masters};    
  \node at (5,-3.9){\includegraphics[scale=0.15]{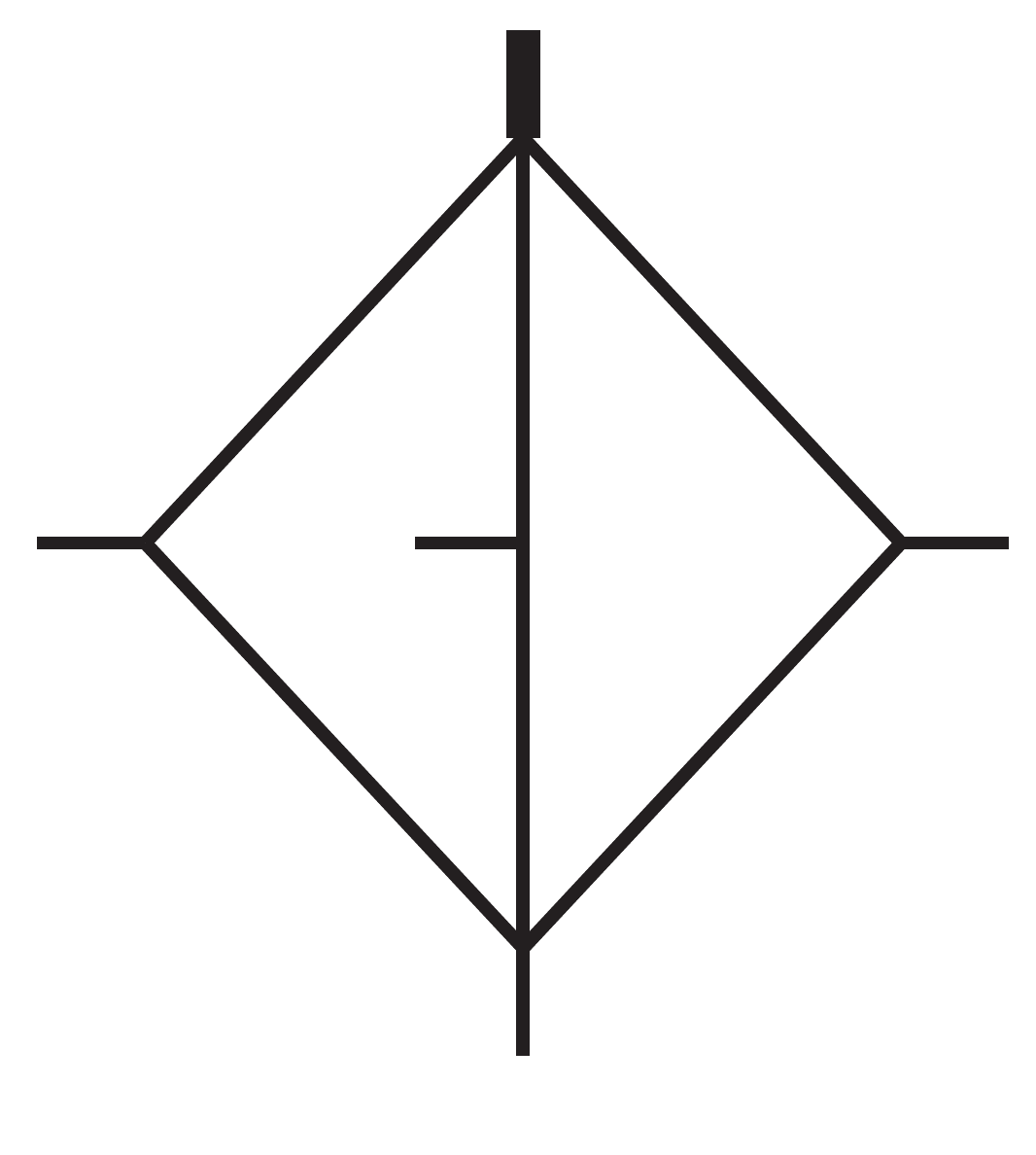}};
	\node at (5,-4.9){1 master};
  \node at (8,-3.9){\includegraphics[scale=0.15]{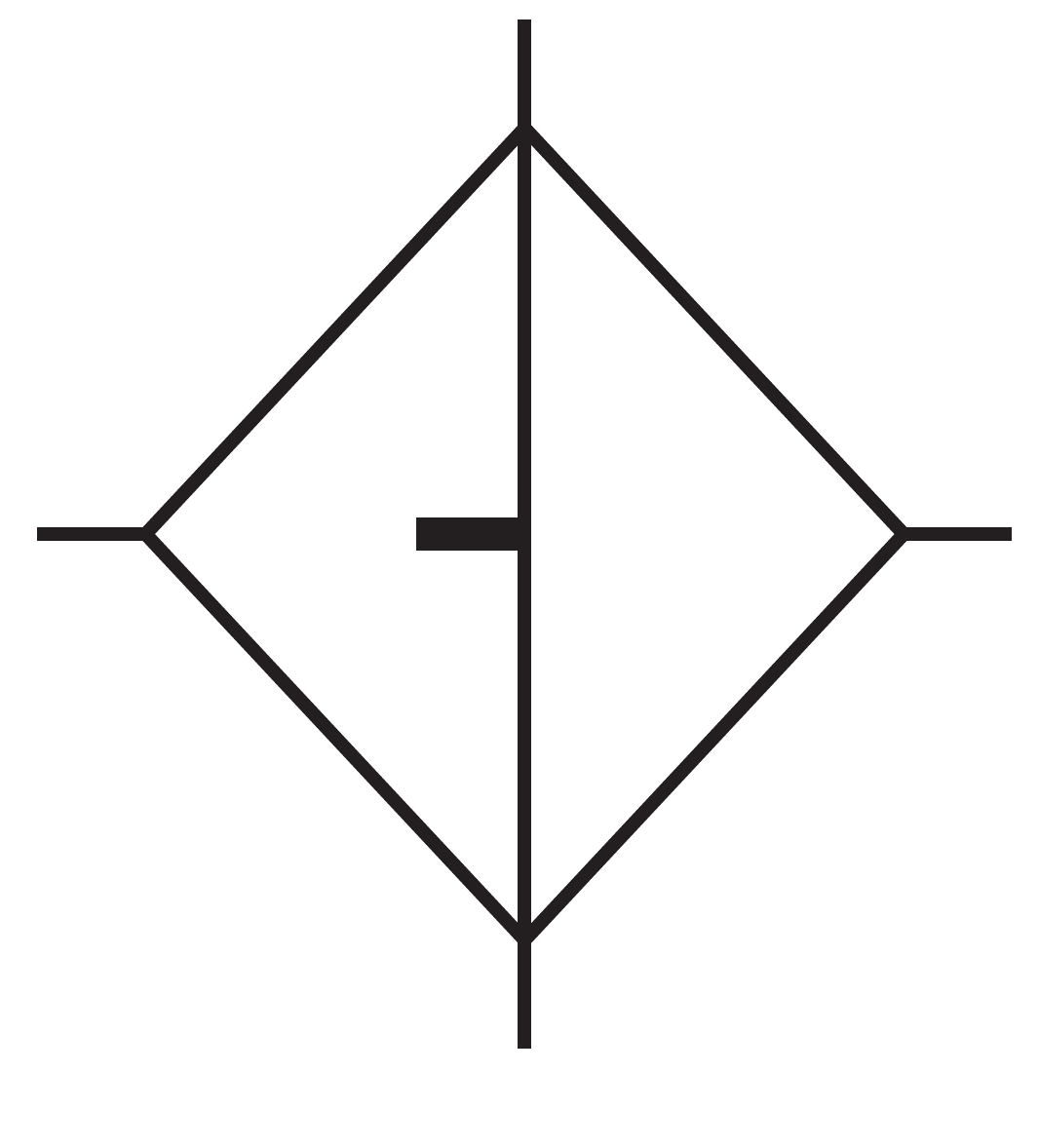}};
	\node at (8,-4.9){4 masters};
  \end{tikzpicture}
  \caption{Propagator structures of two-loop five-point non-planar master integrals in 
  the hexa-box topologies.} 
  \label{fig_master_int} 
\end{figure}
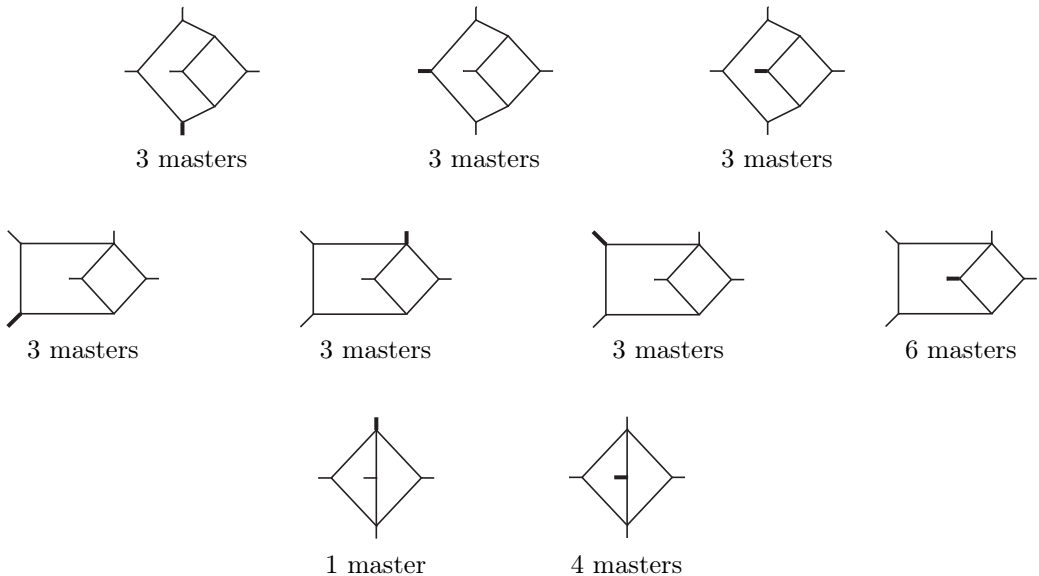

Even though the dimensions given in eq.~\eqref{eq:masterCount} are rather
large, there is a substantial overlap between these different 
spaces, as the same master integrals can appear in different topologies.
Furthermore, some of the master integrals have been computed previously: 
the planar five-point integrals were given in ref.~\cite{Abreu:2020jxa},
and the integrals associated with Feynman diagrams with four 
external legs in 
refs.~\cite{Henn:2014lfa,Caola:2014lpa, Gehrmann:2015ora}. The master integrals
that appear for the first time in the three non-planar hexa-box topologies are depicted
in \fig{fig_master_int}. Finally, we  note that a full set of master integrals
for topology $I^{[\mzz]}_{\rm hb}[\vec \nu]$ has already been computed 
in ref.~\cite{Papadopoulos:2019iam}.


\section{Canonical Differential Equation and Pure Basis}
\label{sec:pureBasisConstruction}

\subsection{Canonical Differential Equation}\label{sec:canDiffEq}

A particularly powerful method to evaluate Feynman integrals with many scales is
by solving the differential equations they satisfy~\cite{Kotikov:1990kg,
Kotikov:1991pm, Bern:1993kr, Remiddi:1997ny, Gehrmann:1999as, Henn:2013pwa}.
Let ${\bf I}$ be a vector containing the set of master integrals associated with a
given topology (such as the $\mzz$, $\zmz$ or $\zzz$ hexa-boxes).
We consider the master integrals as functions of the Mandelstam variables 
$\vec s$, and treat the dimensional regulator $\epsilon=(4-D)/2$ as a 
parameter. The vector ${\bf I}$ satisfies the differential equation
\begin{equation}
\label{eqn:DE}
{\rm d} {\bf I} = \, \overline{\bf M} \,{\bf I}\,,
\end{equation}
where the connection $\overline{\bf M}$ is a matrix of differential 
forms which depends  rationally on the dimensional regulator 
$\epsilon$. The form of the differential equation (\ref{eqn:DE}) follows 
from the fact that differential operators
generate linear combinations of elements in 
the space $Y^{[f]}$ associated with each topology, which can then be mapped 
back into the basis ${\bf I}$. For complicated enough integrals, this procedure 
can be a bottleneck in determining the differential equations.

The construction of the differential equation can be simplified in two ways. 
First, it is clear that the form of the connection $\overline{\bf M}$ 
depends on the basis ${\bf I}$, 
and $\overline{\bf M}$ will be particularly simple if a basis of
so-called `pure' functions is chosen~\cite{Henn:2013pwa}.
For such a basis, the connection can be written as
\begin{equation}
\label{epsFactorizedDE}
\overline{\bf M}=\epsilon\,{\bf M}\,,\qquad
{\bf M} = \sum_{\alpha}\,M_\alpha \,{\rm d}\log{(W_\alpha)} \,,
\end{equation}
where the elements of the matrices $M_\alpha$ are rational numbers and the 
$W_\alpha$ are algebraic functions of the Mandelstam variables $\vec s$, 
known as the `letters' of the `(symbol) alphabet' $\mathcal{A}$ associated 
with the topology under consideration. 
If the connection $\overline{\bf M}$ takes the form given in 
\cref{epsFactorizedDE}, the differential equation is said to be
in \emph{canonical form}.

The second simplification is in the way the connection ${\bf M}$ is
determined once a pure basis has been found 
\cite{Abreu:2018rcw,Abreu:2018aqd,Abreu:2020jxa}. 
Instead of performing fully analytic IBP reductions for the whole system, 
we determine the new letters from much simpler `cut' differential equations, 
where all integrals that do not contain the set of cut propagators are 
set to zero. Once all the letters $W_\alpha$ have been determined, 
the matrices $M_\alpha$ are computed from fully numerical IBP 
reductions.

In the remainder of this section, we will discuss the construction of
the pure bases of master integrals for the non-planar five-point one-mass 
integrals of \cref{fig_master_int}. The determination of the associated
connections ${\bf M}$ will be left to the next section. 
In both cases we will find it useful to consider a `random-direction 
differential equation'~\cite{Abreu:2020jxa}, which allows us to
replace the connection ${\bf M}$ by an algebraic
function of the kinematics and dimensional regulator. More explicitly,
we choose an arbitrary direction $\vec c$ in the six-dimensional space of 
the Mandelstam variables and compute
\begin{equation}
   \vec{c} \cdot \nabla_{\vec{s}}\, {\bf I} 
  = C(\epsilon, \vec{s}\,)
{\bf I}\,,
\label{eq:randomDirectionDE}
\end{equation}
where $\nabla_{\vec{s}}=\{ \frac{\partial}{\partial \offShellScale} , \frac{\partial}{\partial s_{12} }, …\}$ is 
the gradient operator with respect to the Mandelstam variables in \cref{eqn:orderedInvariants}.
For a pure basis, i.e., in the case of a canonical differential equation, 
$C(\epsilon, \vec{s}\,)$ is given by
\begin{equation}
  C(\epsilon, \vec{s}\,) =
      \epsilon\, \sum_{\alpha} M_\alpha \,\vec{c} \cdot 
        \nabla_{\vec{s}}\, \log(W_\alpha)\,,
\label{eq:CAnsatz}
\end{equation}
and, assuming the vector $\vec c$ is generic, it captures the dependence of
${\bf M}$ on the $M_\alpha$ and $W_\alpha$.
In practice, we find the matrix $C(\epsilon, \vec{s}\,)$ particularly useful as it
can be evaluated on numerical kinematic configurations, by performing numerical 
IBP reductions of the left-hand side of \cref{eq:randomDirectionDE}.

\subsection{Pure Basis}

The construction of a pure basis for multi-scale Feynman integrals 
remains challenging, despite much recent progress 
\cite{Lee:2014ioa,Prausa:2017ltv,Gituliar:2017vzm,Meyer:2017joq,Abreu:2018rcw,Wasser:2018qvj,
Abreu:2018aqd,Chicherin:2018old,Dlapa:2020cwj,Henn:2020lye,Abreu:2020jxa}.
Moreover, starting from five external legs, four-dimensional analyzes are often not sufficient,
see e.g.~refs.~\cite{Abreu:2018aqd,Chicherin:2018old}.
In this section we discuss how a basis of pure master integrals for all
the integrals in \cref{fig_master_int} was constructed. As we explain below,
this process requires the calculation of numerical IBP reductions. These were
obtained with two publicly available codes, 
\texttt{Kira 1.2}~\cite{Maierhoefer:2017hyi} 
and \texttt{FIRE6}~\cite{Smirnov:2019qkx}.

Throughout this section, we will often use the functions $\mu_{ij}$
when constructing pure integrals.
These correspond to contractions of 
the components of the loop momenta beyond four dimensions, which
we denote $\ell^{(D-4)}_i$.
Explicitly,
\begin{eqnarray}
\label{eq:mu}
\mu_{ij}= \ell^{(D-4)}_i \cdot \ell^{(D-4)}_j\,.
\end{eqnarray} 
These functions can also be written as polynomials in the 
$\rho_{i,f}$ and the Mandelstam variables $\vec{s}$, 
see~\cref{eqn:orderedInvariants,eq:propvars}. The latter
representation is more convenient if one wants to rewrite integrals
defined with the help of these functions as members of the vector
spaces $Y^{[f]}$. For convenience and to remove any ambiguity related
to conventions, we provide a routine to compute the functions
$\mu_{11}$, $\mu_{22}$ and $\mu_{12}$
in the ancillary file \texttt{anc/determinants.m} .

The integrals listed in  \cref{fig_master_int} can be grouped into two classes:
the cases for which the number of master integrals is the same as in the massless 
limit $p_1^2\rightarrow 0$, and the cases with an increased number of master integrals.
For instance, for the three independent hexa-box topologies on the top row of
\cref{fig_master_int}  we find that the number of master
integrals is the same. In such cases, the trivial generalization of the pure basis
from the massless case \cite{Abreu:2018rcw} to the massive case gives a complete set of
pure master integrals. Despite not being the main difficulty in constructing a complete
basis of master integrals, we note that the pure bases we present here
for these integrals (which we list below) are particularly compact.

\begin{figure}[]
\centering
	\includegraphics[scale=0.31]{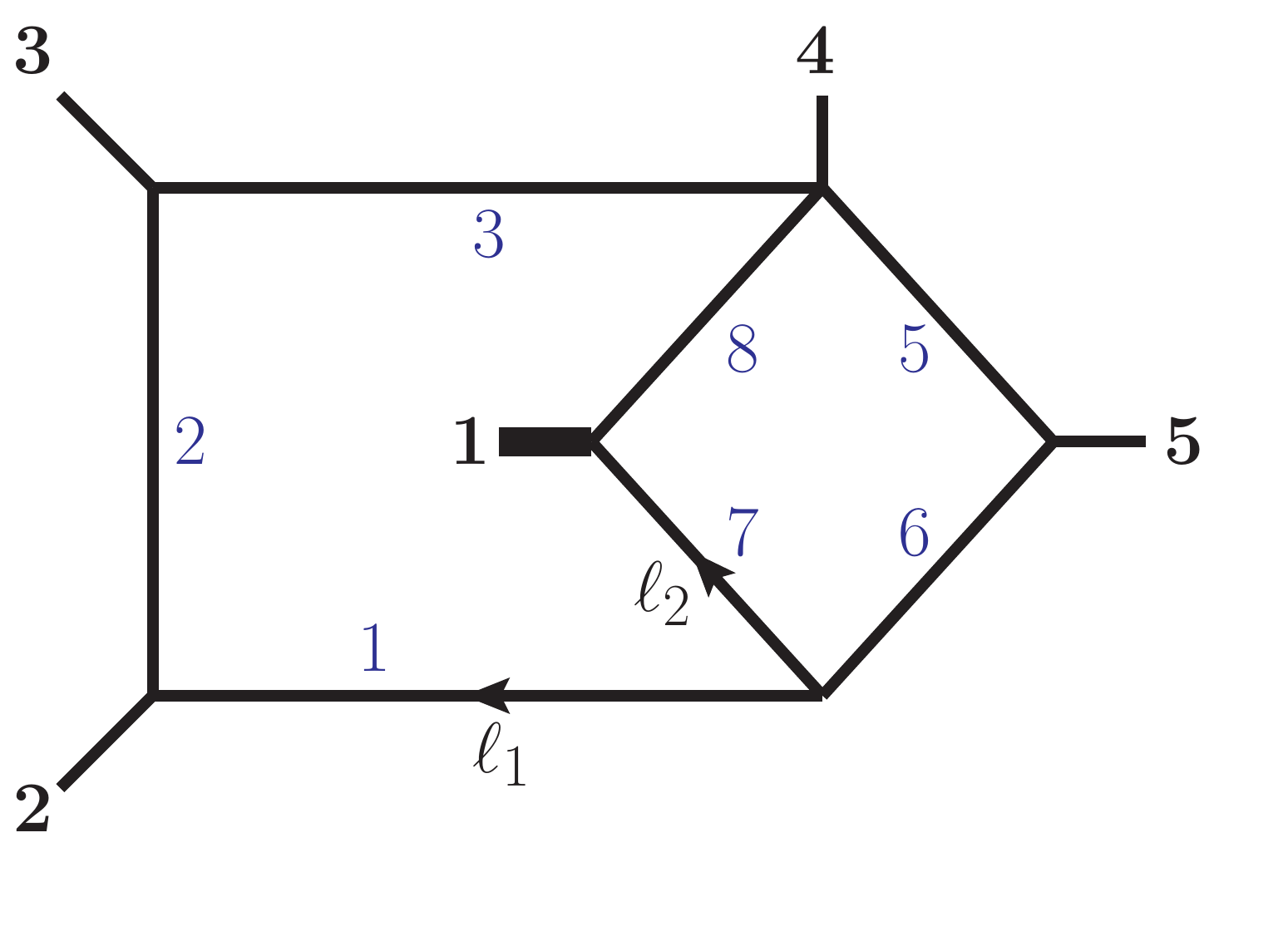}
\caption{Penta-box with 6 master integrals.} 
\label{fig:pentabox}
\end{figure}

\begin{figure}[]
\centering
\includegraphics[scale=0.33]{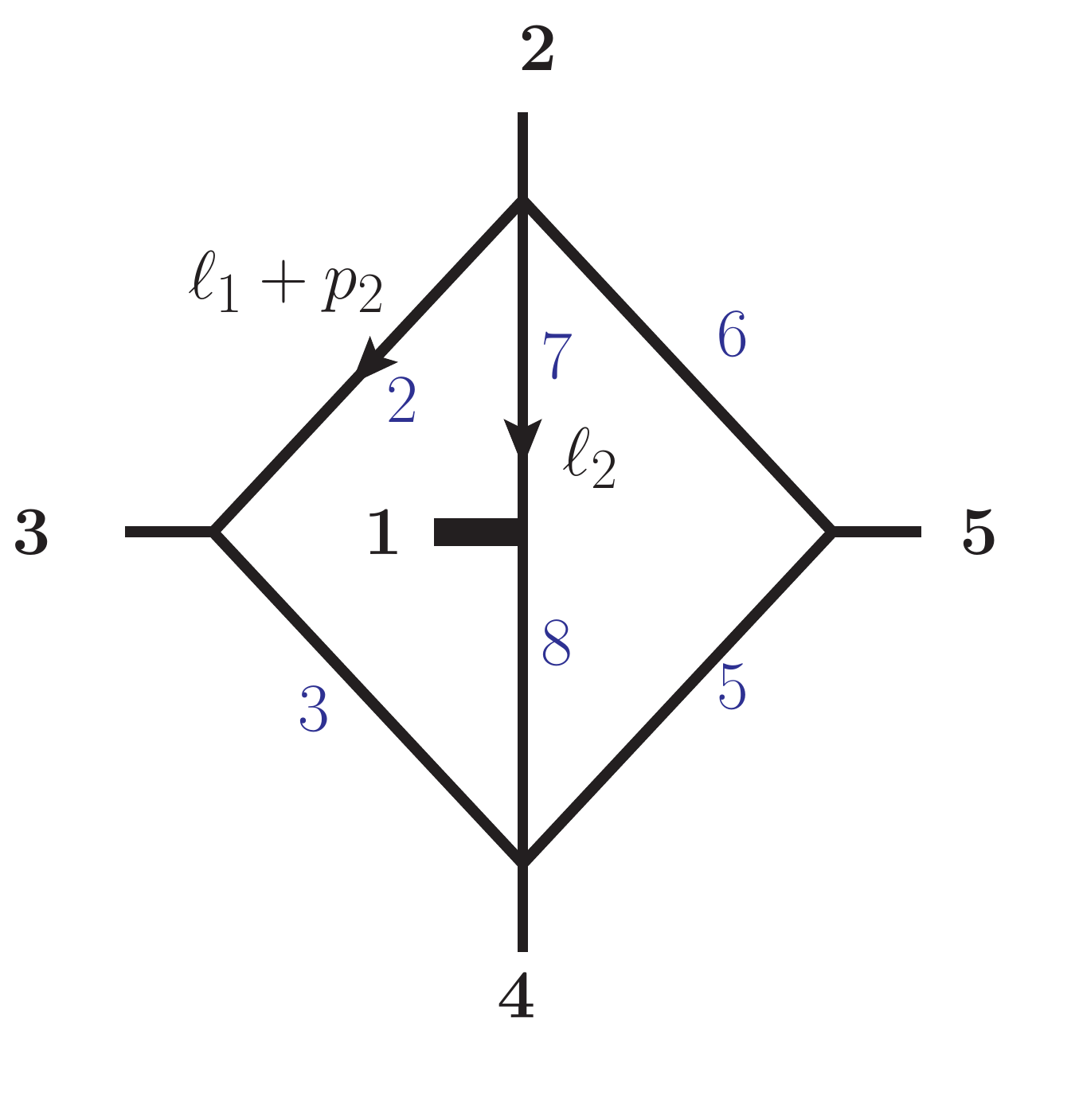}
\caption{Double-box with 4 master integrals.}
\label{fig:doublebox}
 \end{figure}

The cases for which the master-integral count changes are more challenging to handle.
These cases are the penta-box diagram depicted in \cref{fig:pentabox} and the double-box integral 
shown in \cref{fig:doublebox}. 
In order to determine a basis of
pure integrals (beyond those corresponding to generalizations of the massless case) we use a similar approach in both cases.
We start by constructing educated guesses for pure master integrals by studying their
leading singularities \cite{Cachazo:2008vp}, the goal being to check if they can be written as
$d {\rm log}$ forms with unit leading singularity (see e.g.~\cite{Henn:2014qga}).
In complicated cases, this analysis can be done under the condition that all of
the inverse propagators are set to zero (i.e., on their `maximal cut').
Working on the maximal cut has another benefit:
the differential equations on the maximal cut are much simpler, as all integrals that do not involve all 
the cut propagators are set to zero. 
We thus verify if, under these cut conditions, we obtain a canonical
differential equation
of the form of \cref{epsFactorizedDE} or \cref{eq:CAnsatz}. This can
be done fully numerically, by setting the Mandelstam variables and the dimensional regulator
to generic numerical values. In fact, to be more precise, 
at this stage we simply check that the $\epsilon$ dependence factorizes
in the matrix $C(\epsilon,\vec s\,)$ of \cref{eq:CAnsatz}, and  we cannot yet guarantee that
it only involves $d {\rm log}$ forms and matrices of rational numbers. 
Nevertheless, for now we assume that $\epsilon$ factorization implies purity of the basis we 
are constructing (in the next section we discuss how this assumption is verified).
Once we have built enough pure master integrals on the maximal cut, we start releasing cut conditions
and check, numerically, if the $\epsilon$ dependence still factorizes in the differential equation.
If it is not the case, we correct the candidate pure integral with terms proportional to the propagator which is no longer
set to zero. 

Either by generalizing the pure integrals from $p_1^2=0$ to $p_1^2\neq0$ or 
by following the steps above, we constructed bases of master integrals for all the diagrams
depicted in \cref{fig_master_int}. Before listing the integrands corresponding to the pure bases,
let us discuss a particular example in more detail, as it will illuminate the appearance of the
(square-root of the) polynomial in \cref{eq:newRoot}. We consider the double-box integral 
of~\cref{fig:doublebox},
\begin{align}\label{eq:intNorm}
I_{\rm db} = \int \frac{{\rm d}^D\ell_1}{i\pi^{D/2}}
\frac{{\rm d}^D\ell_2 }{i\pi^{D/2}} 
\frac{\mathcal{N}^{(1)}}
{(\ell_1+p_2)^2(\ell_1+p_2+p_3)^2(\ell_1+\ell_2-p_5)^2(\ell_1+\ell_2)^2\ell_2^2(\ell_2+p_1)^2}\,.
\end{align}
By analogy with the planar case \cite{Henn:2013pwa,Henn:2014qga}, given that it is a double-box integral 
we expect that the integral should be pure for some function $\mathcal{N}^{(1)}$ that depends only
on the external kinematics and, furthermore, that it should be possible to compute
this function by studying the leading singularity of the scalar integral in exactly four dimensions.
In order to do this, we consider a loop-by-loop approach and write
\begin{align}
I_{\rm db} \sim \mathcal{N}^{(1)} \int 
\frac{{\rm d}^4\ell_2}{\ell_2^2(\ell_2+p_1)^2}
\left(\int\frac{{\rm d}^4\ell_1 }
{(\ell_1+p_2)^2(\ell_1+p_2+p_3)^2(\ell_1+\ell_2-p_5)^2(\ell_1+\ell_2)^2}\right)\,,
\end{align}
where we did not keep track of conventional numerical normalization factors.
The innermost integral over the loop momentum $\ell_1$ corresponds to a so-called `two-mass easy' box
in four dimensions. It is well know that such an integral has a 
$d {\rm log}$ representation \cite{Arkani-Hamed:2016byb}, and that its leading singularity 
is~(see e.g.~\cite{Abreu:2017ptx})
\begin{align}
{\cal R}_{\rm box}&=\frac{1}{(\ell_2-p_2-p_3)^2  (\ell_2+ p_1+p_3+p_4 )^2 - (\ell_2-p_2)^2 (\ell_2+p_1+p_4)^2}.
\end{align}
To proceed, we use the factorization properties of this leading singularity 
\cite{Bern:2014kca}, i.e.~we define momenta $q$ and $\bar{q}$ such that
\begin{equation}
q\cdot (\ell_2-p_2)  = \langle p_3 | (\slash{\hspace{-.17cm}\ell}_2-\slash{\hspace{-.19cm}p}_2) | p_5 ] 
\quad \textrm{and} \quad 
\bar{q}\cdot (\ell_2-p_2)  = [p_3 | (\slash{\hspace{-.17cm}\ell}_2-\slash{\hspace{-.19cm}p}_2) | p_5 \rangle,
\end{equation}
which allow us to factorize the denominator of ${\cal R}_{\rm box}$,
\begin{align}
{\cal R}_{\rm box}&=\frac{1}{(q\cdot [\ell_2-p_2]) \, (\bar q\cdot [\ell_2-p_2])} .
\end{align}
We are now left with the task of determining $\mathcal{N}^{(1)}$ such that the differential form
\begin{equation}
  \omega=\frac{\mathcal{N}^{(1)}\,{\rm d}^{4}\ell_2}
  { (\ell_2)^2 (\ell_2 +p_1)^2 (q\cdot [\ell_2-p_2]) \, (\bar q \cdot [\ell_2-p_2])}
  \label{eq:targetForm}
\end{equation}
has unit leading singularity.
This can be achieved in a simple and enlightening manner by using
the embedding-space formalism of ref.~\cite{SimmonsDuffin:2012uy},
see also refs.~\cite{Caron-Huot:2014lda,Abreu:2017ptx} for applications
to Feynman integrals. 

This formalism upgrades
the 4-dimensional form to a form in a 6-dimensional projective space which is
integrated over the light cone.
We label points in the 6-dimensional embedding space with capital letters. A
vector $V$ in embedding space is written in terms of a
four-dimensional vector $v$ and two light-cone components, i.e.,
\begin{equation}
    V = (v, V^+, V^-)\,.
\end{equation}
The scalar product between two embedding-space vectors $V$ and $W$ is defined as
\begin{equation}
  \label{eq:EmbeddingScalarProductDefinition}
(V W) = -2 (v \cdot w) - V^+ W^- - V^- W^+\,.
\end{equation}
We then introduce the following vectors in embedding space:
\begin{align}
  \begin{split}
    Y &= (\ell, -\ell^2, 1), \qquad
    X_0 = (0, 0, 1), \qquad
    X_1 = (-p_1, -p_1^2, 1), \\
    X_q &= \left(-\frac{1}{2}q, q \cdot p_2, 0\right), \qquad
    X_{\bar{q}} = \left(-\frac{1}{2}\bar{q}, \bar{q} \cdot p_2, 0\right).
    \end{split}
\end{align}
Note that as $q$ and $\bar{q}$ are massless, all of these vectors square to zero
with the scalar product of~\eqref{eq:EmbeddingScalarProductDefinition}.
The $Y$ vector corresponds to the loop momentum, and the $X_i$ vectors
correspond to the external momenta in each denominator.
In these variables, the differential form in \eqref{eq:targetForm} can be rewritten as
\begin{equation}
\omega=\frac{{\rm d}^6Y \delta\left( (YY) \right)}{\textrm{vol}(\textrm{GL}(1))}
\frac{\mathcal{N}^{(1)}}{(Y X_0) (Y X_1) (Y X_q) (Y X_{\bar{q}})},
\end{equation}
which can be analyzed in exactly the same way as the integrands of one-loop integrals
were studied in ref.~\cite{Abreu:2017ptx}. In particular, to compute the leading
singularity of this integral we simply need to change variables to the propagators, 
which are now linear in $Y$, and then impose the conditions 
$(Y X_I)=0$, for $I=0,1,q, \bar{q}$. Since $\omega$ only has simple poles,
this amounts to computing the Jacobian determinant of the change of variables and imposing
the conditions $(Y X_I)=0$. The leading singularity of $\omega$ is then
\begin{equation}
	\mathcal{R}_\omega=\frac{\mathcal{N}^{(1)}}{\sqrt{\nsqrt^{(2)}}}\,,
\end{equation}
where
\begin{align}
\begin{split}
\nsqrt^{(2)} &=
\mathrm{det}\left(
    \begin{array}{cccc}
(X_0 X_0) &(X_0 X_1)&(X_0 X_q)&(X_0 X_{\bar q})\\
(X_1 X_0) &(X_1 X_1)&(X_1 X_q)&(X_1 X_{\bar q})\\
(X_q X_0) &(X_q X_1)&(X_q X_q)&(X_q X_{\bar q})\\
(X_{\bar q} X_0) &(X_{\bar q} X_1)&(X_{\bar q} X_q)&(X_{\bar q} X_{\bar q})
    \end{array}
\right), \\
&=
  \mathrm{det}\left(
\begin{array}{cccc}
0 & p_1^2 & -q \cdot p_2 & - \bar q \cdot p_2 \\
p_1^2 & 0 & q \cdot p_4 & \bar q \cdot p_4 \\
-q \cdot p_2 & q \cdot p_4 & 0 & -\frac{1}{2} q \cdot \bar q \\
-\bar q \cdot p_2 & \bar q \cdot p_4 & - \frac{1}{2} q \cdot \bar q & 0
\end{array}
  \right)\,,\\
&=( {\offShellScale} ( {s_{12}}- {s_{45}})- {s_{12}} ( {s_{15}}+ {s_{23}})+
 {s_{15}} {s_{45}}+ {s_{23}}  {s_{34}}- {s_{34}}  {s_{45}})^2\\
 &+4  {s_{12}}  {s_{23}}( {\offShellScale}- {s_{15}}) ( {s_{12}}- {s_{34}}- {s_{45}})\,.
\end{split}
\label{eq:SigmaDeterminant}
\end{align}
This expression is a permutation of the polynomial $\nsqrt$ defined in \cref{eq:newRoot},
corresponding to $p_4\leftrightarrow p_5$.
To obtain the representation of $\nsqrt^{(2)}$ in terms of Mandelstam invariants, 
we used that
\begin{align}
    q \cdot \bar{q} &= - s_{35}\,,\\
    (q \cdot p_a) (\bar{q} \cdot p_b) &= \mathrm{tr}_+\left(3, a, 5, b\right)\,,
\end{align}
where 
\begin{equation}\label{eq:trpm}
  \mathrm{tr}_{\pm}(i_1 \ldots i_n) = 
  \mathrm{tr}\left(\left[ \frac{1 \pm \gamma_5}{2} \right] \slashed{p}_{i_1}
   \cdots \slashed{p}_{i_n}\right)\,,
\end{equation}
which can be readily written in terms of Mandelstam
invariants (see e.g.~eq.~(5.10) of ref.~\cite{Abreu:2020jxa}).

In summary, given the above calculation we expect that the integral
\begin{align}
I_{\rm db} = \int \frac{{\rm d}^D\ell_1}{i\pi^{D/2}}
\frac{{\rm d}^D\ell_2 }{i\pi^{D/2}} 
\frac{\sqrt{\nsqrt^{(2)}}}
{(\ell_1+p_2)^2(\ell_1+p_2+p_3)^2(\ell_1+\ell_2-p_5)^2(\ell_1+\ell_2)^2\ell_2^2(\ell_2+p_1)^2}
\end{align}
should be pure. Through similar arguments, we can build pure candidates for
all the master integrals required for the diagrams in \cref{fig_master_int}, and
verify numerically that they satisfy a differential equation where the $\epsilon$
dependence factorizes from the connection. We close this section by listing 
the candidate pure basis we have computed for each of these diagrams.
We note that the same information can be found in the ancillary files
\texttt{anc/f/f\_pure\_basis.m} for $\texttt{f}\in\{\texttt{mzz}, \texttt{zmz}, \texttt{zzz}\}$, 
and we also include a pictorial representation of the bases in
\texttt{anc/f/f\_graphs.m} which was generated using 
ref.~\cite{Georgoudis:2016wff}.

 \subsection*{Hexa-boxes}
\begin{minipage}{0.3\textwidth}
\begin{figure}[H]
\includegraphics[width=4.3cm]{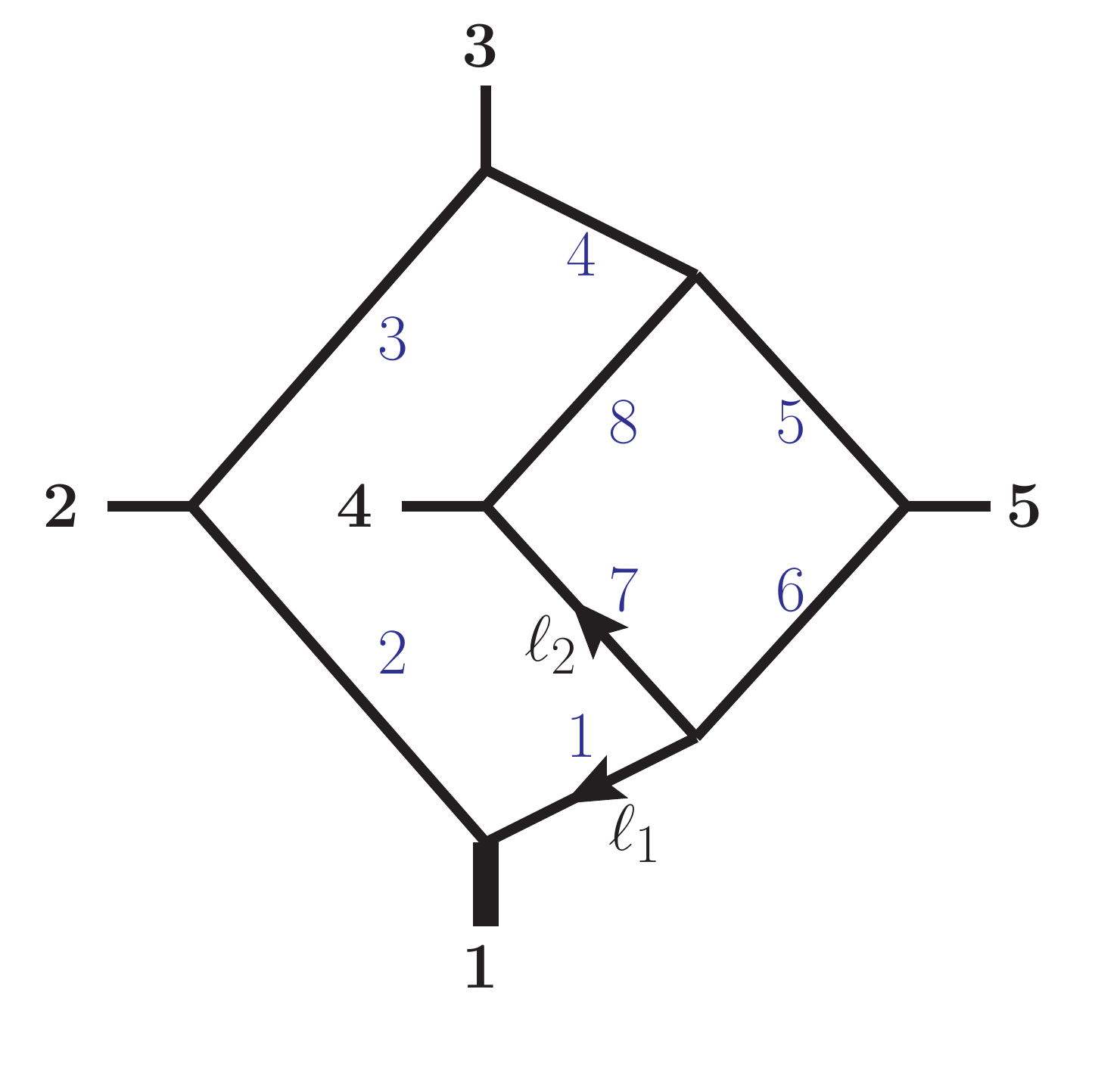}
\end{figure}
 \end{minipage}
\begin{minipage}{0.7\textwidth}
\begin{align}\begin{split}
\mathcal N^{(1)}_{\textrm{hb},\mzz} &= \epsilon^4 \sqrt{\Delta_5}\, (\ell_1-p_4)^2 \mu_{11}\,,\\
\mathcal N^{(2)}_{\textrm{hb},\mzz} &= \epsilon^4 \sqrt{\Delta_5}\,  (\ell_1-p_5)^2 \mu_{11}\,, \\
\mathcal N^{(3)}_{\textrm{hb},\mzz} &= \epsilon^4  s_{12}  s_{23}\left[(\ell_1-p_4)^2(\ell_1-p_5)^2-\rho_1\rho_4\right] \,. \hspace{1.6cm}
\end{split}\end{align}
\end{minipage}\\
\begin{minipage}{0.3\textwidth}
\begin{figure}[H]
\includegraphics[width=4.3cm]{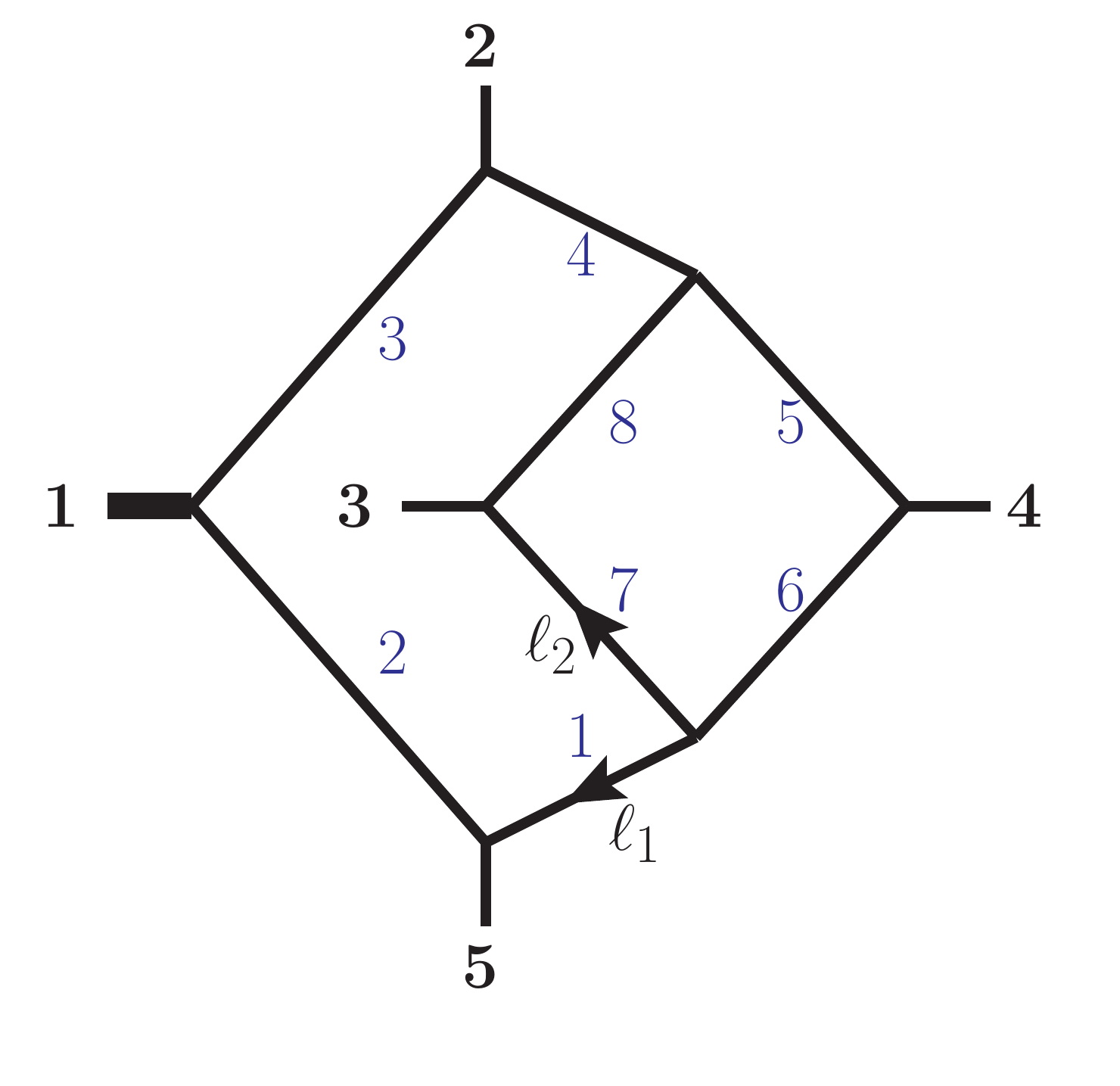}  
\end{figure}
\end{minipage}
\begin{minipage}{0.7\textwidth}
\begin{align}\begin{split}
\mathcal N^{(1)}_{\textrm{hb},\zmz} &= \epsilon^4  \sqrt{\Delta_5}\, (\ell_1-p_3)^2  \mu_{11}\,,\\
\mathcal N^{(2)}_{\textrm{hb},\zmz} &= \epsilon^4  \sqrt{\Delta_5}\, (\ell_1-p_4)^2  \mu_{11}\,,\\
\mathcal N^{(3)}_{\textrm{hb},\zmz} &=\epsilon^4 \left[s_{12}s_{15}-p_1^2s_{34}\right]\left[(\ell_1\!-\!p_3)^2(\ell_1\!-\!p_4)^2-\rho_1\rho_4\right]\,. \hspace{0.4cm}
\end{split}\end{align}
\end{minipage}\\
\begin{minipage}{0.3\textwidth}
\begin{figure}[H]
\includegraphics[width=4.3cm]{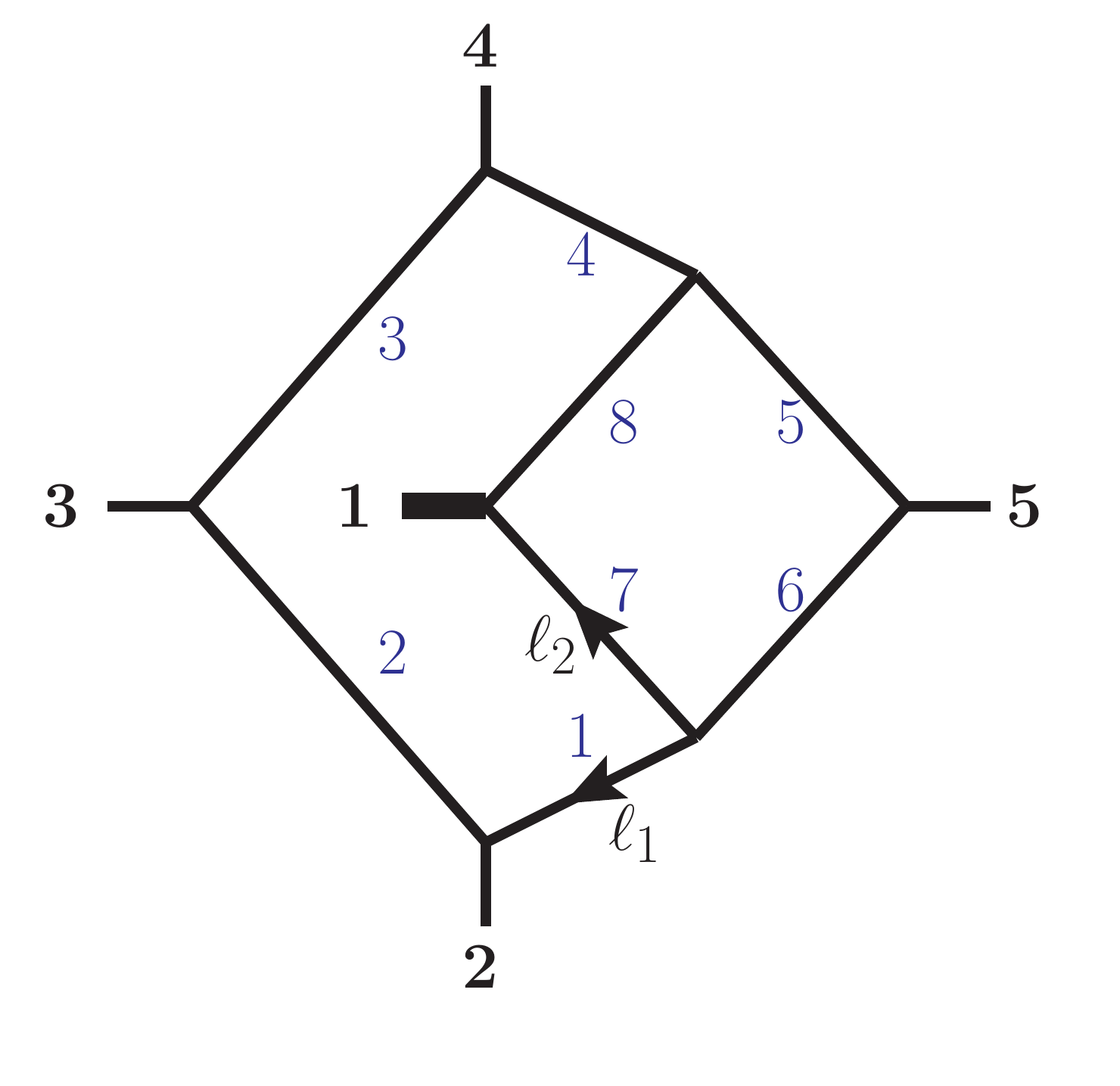}  
\end{figure}
\end{minipage}
\begin{minipage}{0.7\textwidth}
\begin{align}\begin{split}
\mathcal N^{(1)}_{\textrm{hb},\zzz} &= \epsilon^4  \sqrt{\Delta_5}\,  (\ell_1-p_1)^2 \mu_{11}\,,\\
\mathcal N^{(2)}_{\textrm{hb},\zzz} &= \epsilon^4  \sqrt{\Delta_5}\,  (\ell_1-p_5)^2 \mu_{11}\,,\\
\mathcal N^{(3)}_{\textrm{hb},\zzz} &= \epsilon^4s_{23}s_{34}\left[(\ell_1-p_1)^2(\ell_1-p_5)^2-\rho_1\rho_4\right]\,. \hspace{1.9cm}
\end{split}\end{align}
\end{minipage}

\subsection*{Penta-boxes}
\begin{minipage}{0.3\textwidth}
\begin{figure}[H]
\includegraphics[width=4.3cm]{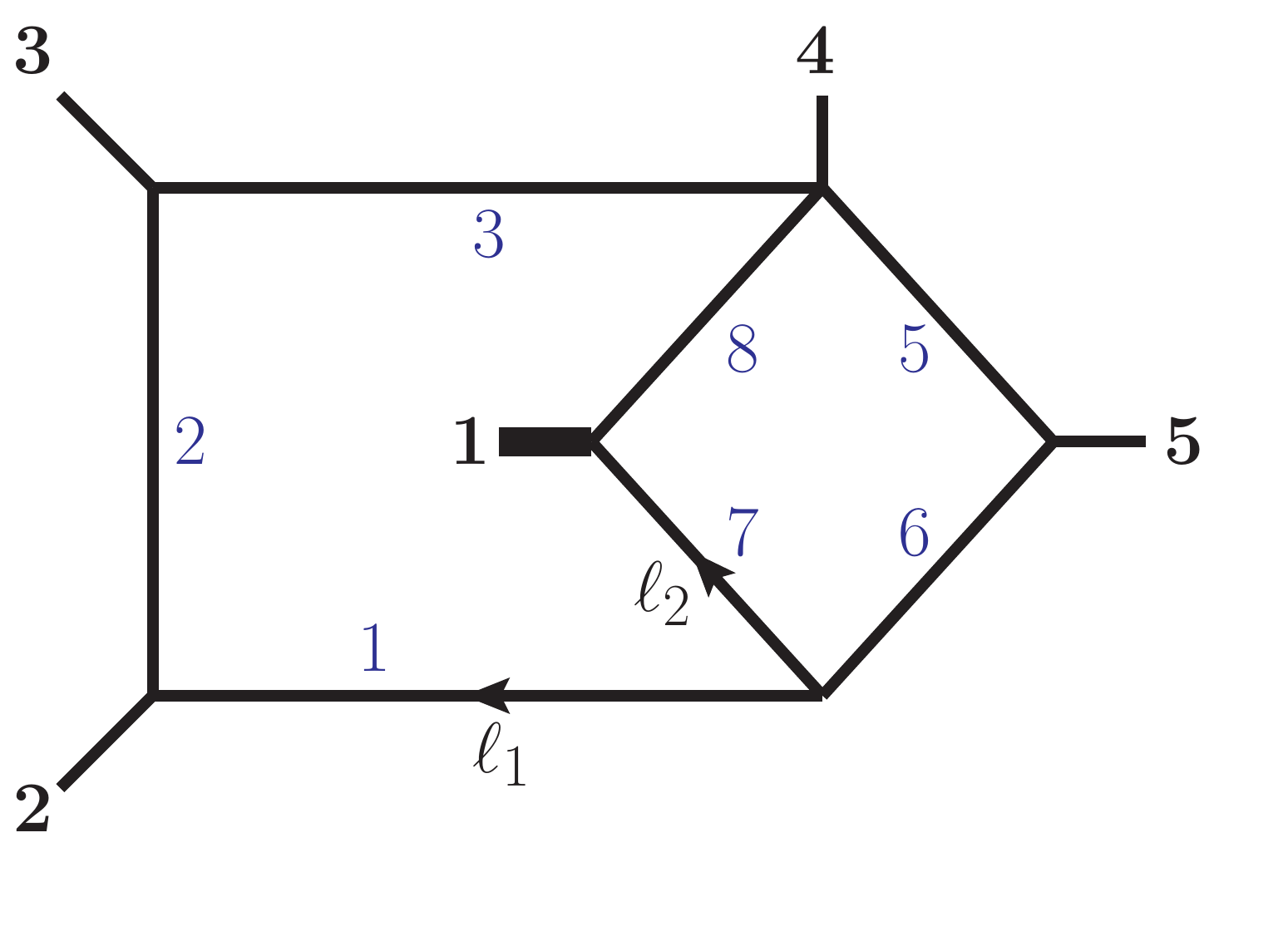}  
\end{figure}
\end{minipage}
\begin{minipage}{0.7\textwidth}
\begin{align}\begin{split}
\mathcal N^{(1)}_{\textrm{pb},1}  &=  \epsilon^4 \left[s_{23}(s_{34}-s_{12}-s_{15}+p_1^2)(\ell_1-p_1)^2 + C_{\textrm{pb},1}^{(1)}\,\rho_1\right]\,,\\
\mathcal N^{(2)}_{\textrm{pb},1}  &=  \epsilon^4 \left[s_{23}s_{12}(\ell_1-p_5)^2+C_{\textrm{pb},1}^{(2)}\,\rho_1\right]\,,\\
\mathcal N^{(3)}_{\textrm{pb},1}  &=\epsilon^4 s_{23}\left[(\ell_1-p_1)^2(\ell_1-p_5)^2-\rho_1(\ell_1-p_1-p_5)^2\right]\,,\\
\mathcal N^{(4)}_{\textrm{pb},1}  &=\epsilon^3  \sqrt{\Delta_5}\, p_1^2 \,\frac{\mu_{12}+\mu_{11}}{\rho_8}\,,\\
\mathcal N^{(5)}_{\textrm{pb},1}  &=\epsilon^4 \sqrt{\Delta_5}\,  \mu_{12}\,,\\
\mathcal N^{(6)}_{\textrm{pb},1}  &=\epsilon^4 \sqrt{\Delta_5}\,  \mu_{11}\,.
\end{split}\end{align}
\end{minipage}\\
\begin{minipage}{0.3\textwidth}
\begin{figure}[H]
\includegraphics[width=4.3cm]{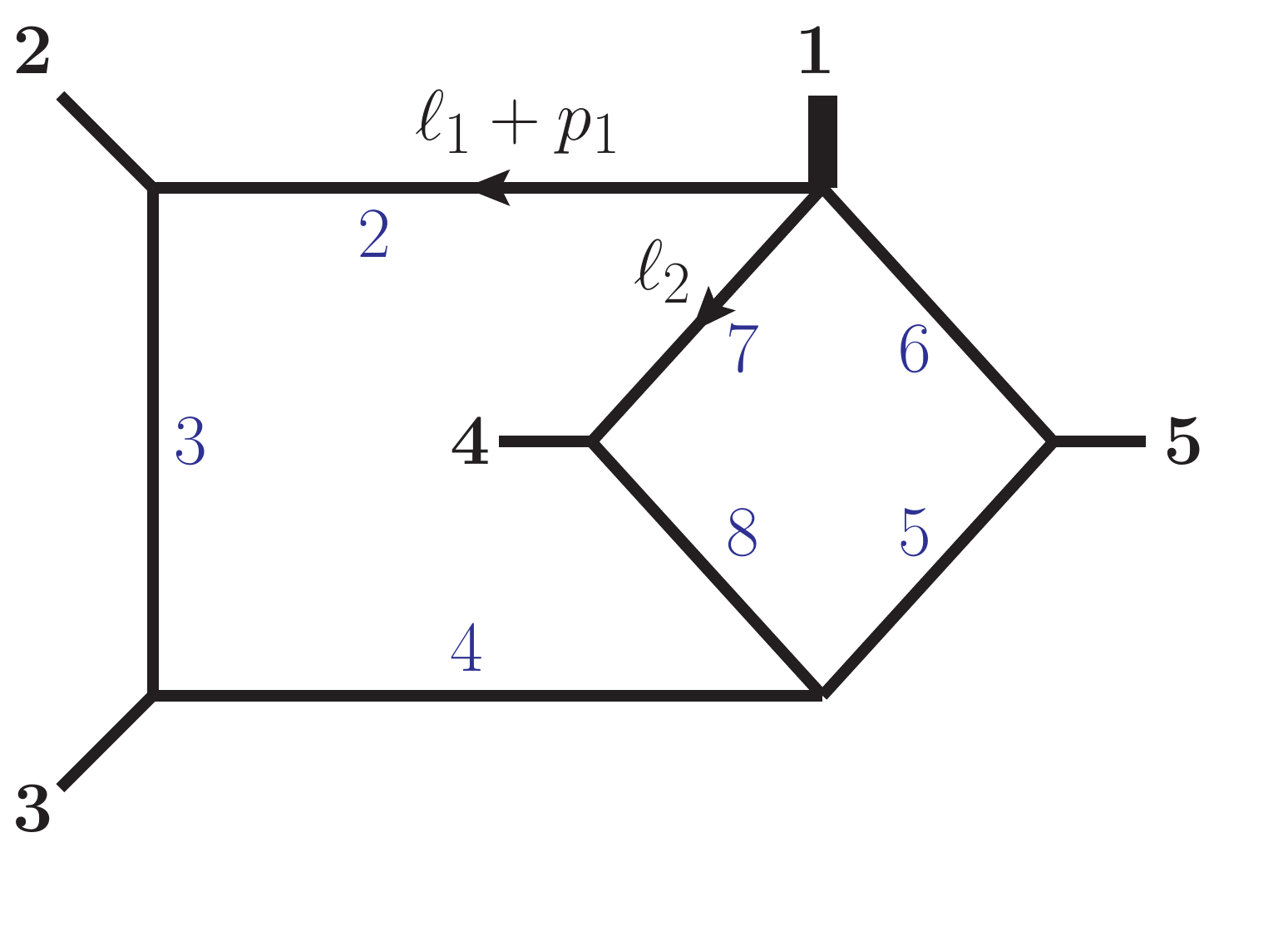}  
\end{figure}
\end{minipage}
\begin{minipage}{0.7\textwidth}
\begin{align}\begin{split}
\mathcal N^{(1)}_{\textrm{pb},2}  &=  \epsilon^4 \left[s_{23}s_{34}(\ell_1-p_4)^2 +C_{\textrm{pb},2}\,\rho_4\right]\,,\\
\mathcal N^{(2)}_{\textrm{pb},2}  &=  \epsilon^4 \sqrt{\Delta_5}\,  \mu_{11}\,,\\
\mathcal N^{(3)}_{\textrm{pb},2}  &=\epsilon^4 s_{23}\left[(\ell_1-p_4)^2(\ell_1-p_5)^2-\ell_1^2\rho_4\right]\,. \hspace{2.3cm}
\end{split}\end{align}
\end{minipage}\\
\begin{minipage}{0.3\textwidth}
\begin{figure}[H]
\centering
\includegraphics[width=4.3cm]{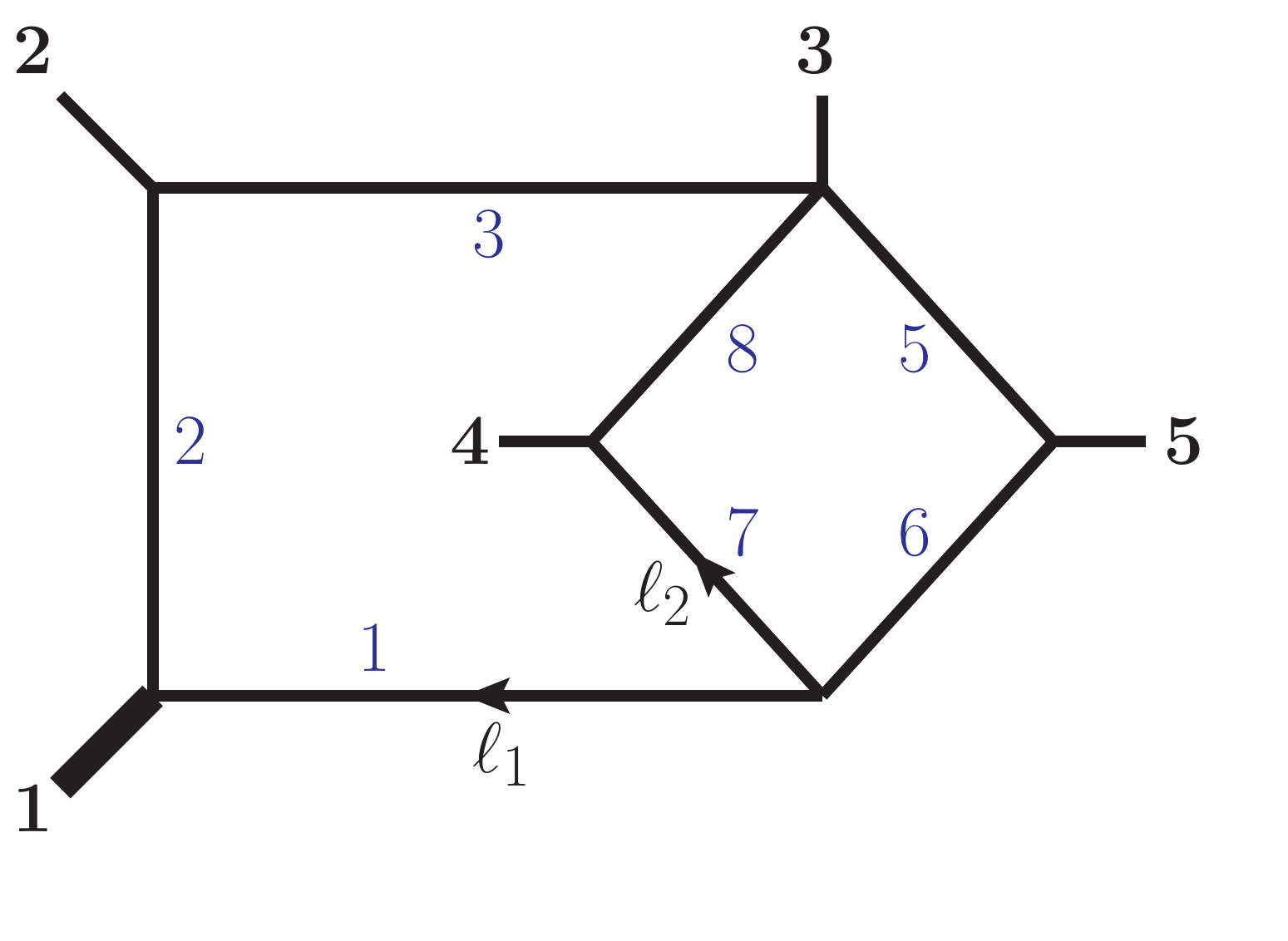} 
\end{figure}
\end{minipage}
\begin{minipage}{0.7\textwidth}
\begin{align}\begin{split}
\mathcal N^{(1)}_{\textrm{pb},3}  &=  \epsilon^4 \left[(s_{12}s_{15}-s_{34}\offShellScale)(\ell_1-p_4)^2 + C_{\textrm{pb},3}\,\rho_1\right]\,,\\
\mathcal N^{(2)}_{\textrm{pb},3}  &=  \epsilon^4 \sqrt{\Delta_5}\, \mu_{11}\,,\\
\mathcal N^{(3)}_{\textrm{pb},3}  &=\epsilon^4 (s_{12}-\offShellScale)\left[(\ell_1\!-\!p_4)^2(\ell_1\!-\!p_5)^2-\rho_1(\ell_1\!-\!p_4\!-\!p_5)^2\right]\,. 
\end{split}\end{align}
\end{minipage}\\
\begin{minipage}{0.3\textwidth}
\begin{figure}[H]
\hspace{0.75cm}
\includegraphics[width=4.3cm]{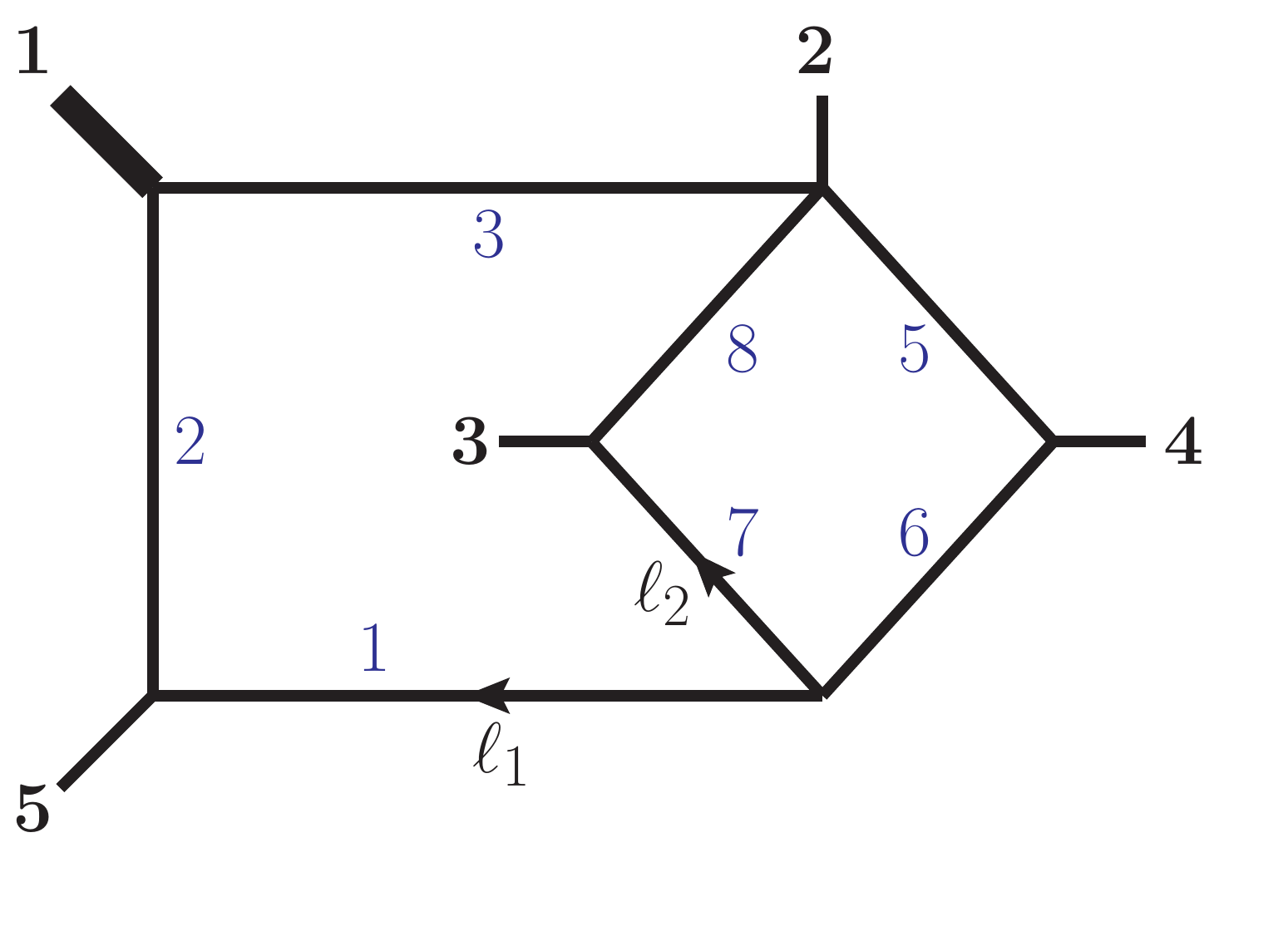}  
\end{figure}
\end{minipage}
\begin{minipage}{0.7\textwidth}
\begin{align}\begin{split}
\mathcal N^{(1)}_{\textrm{pb},4}  &=  \epsilon^4 \left[s_{15}s_{45}(\ell_1-p_3)^2 + C_{\textrm{pb},4}\,\rho_1\right]\,,\\
\mathcal N^{(2)}_{\textrm{pb},4}  &=  \epsilon^4 \sqrt{\Delta_5}\, \mu_{11}\,,\\
\mathcal N^{(3)}_{\textrm{pb},4}  &=\epsilon^4 (s_{15}-\offShellScale)\left[(\ell_1\!-\!p_3)^2(\ell_1\!-\!p_4)^2-\rho_1(\ell_1\!-\!p_3\!-\!p_4)^2\right]\,,
\end{split}\end{align}
\end{minipage}
with
\begin{align}
 \begin{split}
C_{\textrm{pb},1}^{(1)} &= \frac{1}{2}\left(s_{12}(s_{23}-s_{15})+p_1^2(s_{12}-s_{45})  +s_{15} s_{45} -s_{34} (s_{23} + s_{45})\right)\,,\\
C_{\textrm{pb},1}^{(2)} &= -s_{23} s_{34} - C_{\textrm{pb},1}^{(1)} \,, \\
C_{\textrm{pb},2}       &= -\frac{1}{2} \left(s_{12} \left(s_{23}-s_{15}\right) + p_1^2 s_{34}  +s_{23} s_{34}- s_{45}\left (s_{34} -s_{15}\right)\right)\,,\\
C_{\textrm{pb},3}       &= -\frac{1}{2} \left(s_{12} \left(s_{23}+s_{15}\right)-  s_{34}\left( p_1^2+   s_{23}-s_{45}\right) - s_{15} s_{45}\right)\,,\\
C_{\textrm{pb},4}       &= \frac{1}{2}\,\big( s_{12} \left(s_{23}-s_{15}\right) +  s_{34}\left( s_{45}-s_{23}\right)-s_{15}s_{45} \big)\,.
\end{split}
\end{align}\\

\subsection*{Double-boxes}
\begin{minipage}{0.29\textwidth}
\begin{figure}[H]
\includegraphics[width=4.3cm]{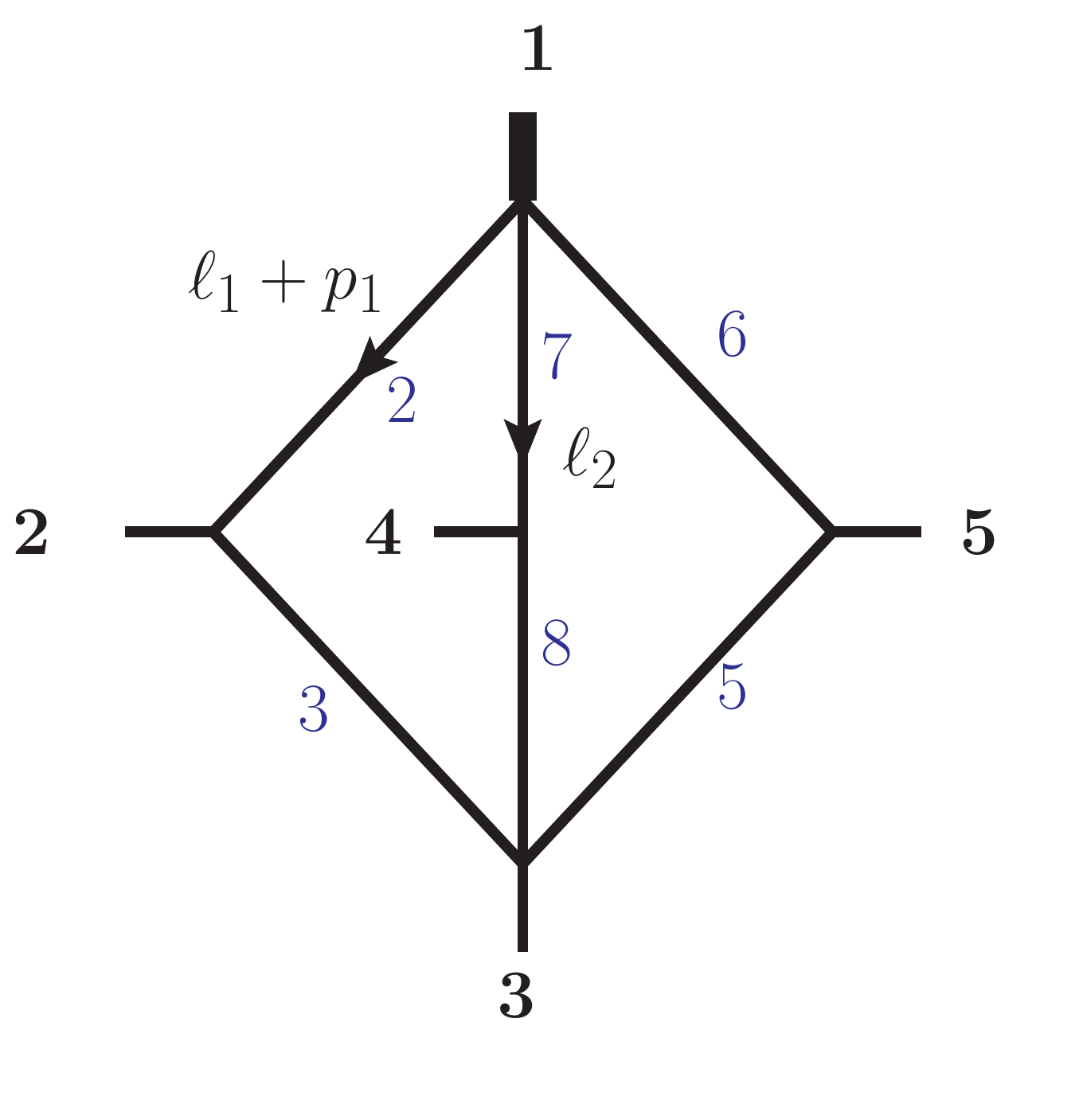}  
\end{figure}
\end{minipage}
\begin{minipage}{0.71\textwidth}
\begin{align}\begin{split}
\mathcal N_{\textrm{db},1}  &=  \epsilon^4  \sqrt{\Delta_5}\,.\hspace{6.2cm}
\end{split}\end{align}
\end{minipage}\\
\begin{minipage}{0.29\textwidth}
\begin{figure}[H]
\includegraphics[width=4.4cm]{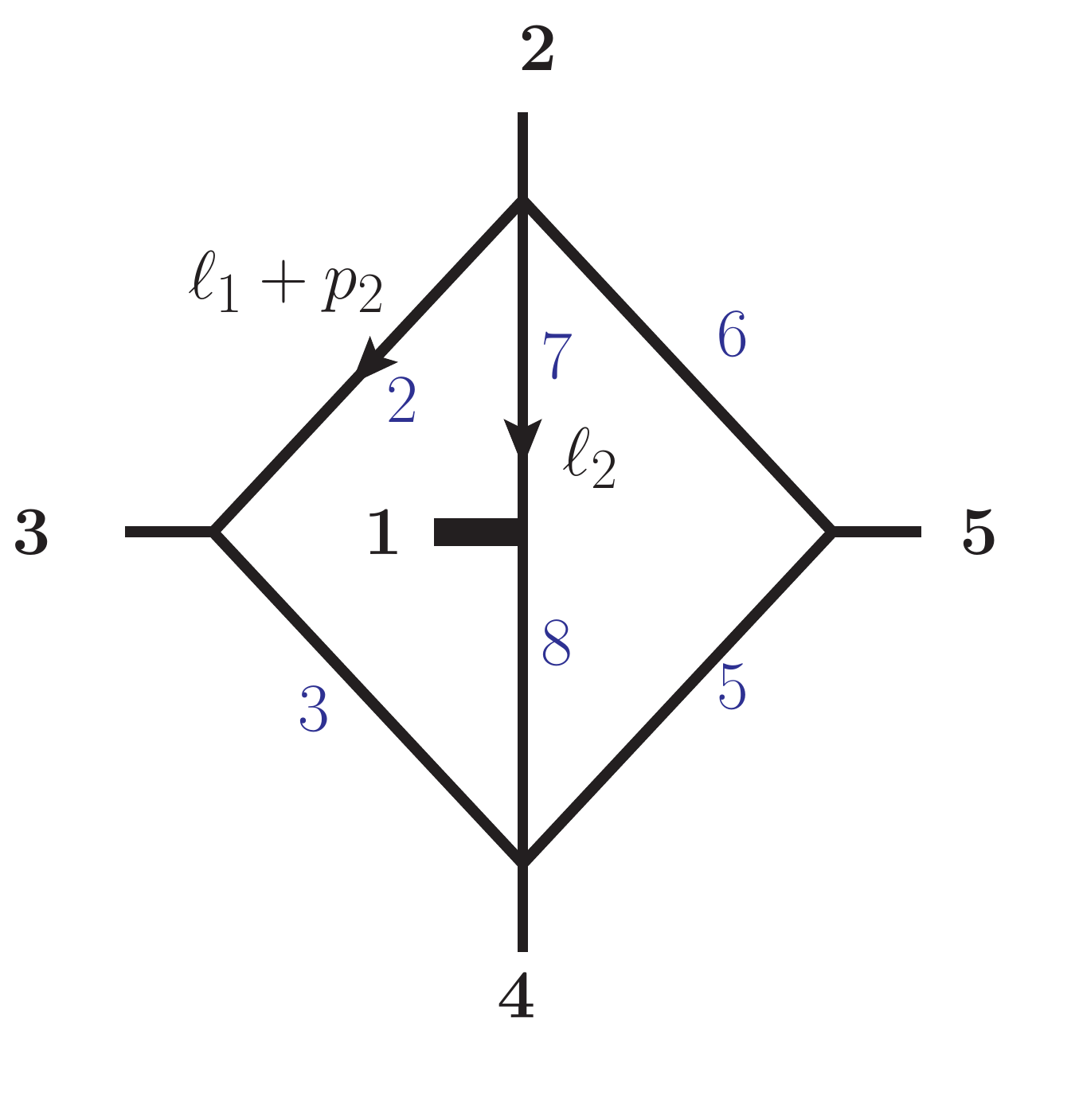}  
\end{figure}
\end{minipage}
\begin{minipage}{0.71\textwidth}
\begin{align}\begin{split}
\mathcal N^{(1)}_{\textrm{db},2}  &=  \epsilon^4 \sqrt{\Sigma_5^{(2)}}\,,\\
\mathcal N^{(2)}_{\textrm{db},2}  &=  \epsilon^4 \frac{1}{8} \left\{ 
		{\rm tr} \left[(\slashed{\ell}_2-\slashed{p}_2) \slashed{p}_3 \slashed{p}_1\slashed{p}_5  \right]  +   
 		{\rm tr}\left[ (\slashed{\ell}_2\!+\!\slashed{p}_1\!+\!\slashed{p}_4) \slashed{p}_3 \slashed{p}_1 \slashed{p}_5  \right] \right.\\
		& \left.+ 8 \left[ (\ell_2+p_1)^2 - \ell_2^2 \right] ( s_{12}-s_{34}-s_{45} )
		\right\}\,,\\
\mathcal N^{(3)}_{\textrm{db},2}  &=\epsilon^3 \sqrt{\Delta_5}\,  \mu_{12}\left(\frac{1}{\rho_7}+\frac{1}{\rho_8}\right),\\
\mathcal N^{(4)}_{\textrm{db},2}  &=\epsilon^3 \left[\frac{\offShellScale}{\rho_7} \left[(\ell_2-p_2-p_3)^2(\ell_2+p_1+p_3+p_4)^2\right.\right. \\
&\left.\left.-(\ell_2-p_2)^2(\ell_2+p_1+p_4)^2\right]+C^{(4)}_{\textrm{db},2}\right]\,,
\end{split}\end{align}
\end{minipage}
with
\begin{align}
 \begin{split}
C^{(4)}_{\textrm{db},2}&= (s_{23}-s_{45}-\offShellScale)\frac{\rho_2 \rho_5}{\rho_3}-(s_{12}+s_{15})\frac{\rho_3\rho_6}{\rho_2}+\frac{1}{\epsilon}(1-2 \epsilon)(1-3 \epsilon)\frac{\offShellScale}{s_{12}-\offShellScale}\rho_3\rho_5\\
&+\frac{1}{\epsilon^2}(1-2 \epsilon)(2-3 \epsilon)(1-3 \epsilon)\frac{\offShellScale}{(s_{12}-\offShellScale)s_{12}}\rho_3\rho_5\rho_7.
\end{split}
\end{align}

\section{Analytic Differential Equations}
\label{sec:DifferentialEquations}

Having constructed bases of master integrals
for which the $\epsilon$ dependence factorizes in the differential equations,
we now construct the analytic form of the connections to verify that they 
are indeed of the form given in \cref{epsFactorizedDE}. We will assume that
$\epsilon$ factorization implies the form of \cref{epsFactorizedDE}, which we 
take as an ansatz. After determining the algebraic functions $W_\alpha$ that
constitute the letters of the alphabet of each topology, we will then fit the 
matrices of rational numbers $M_\alpha$
from numerical evaluations of the differential equations. The success of 
this procedure will both confirm that the bases of master integrals 
introduced in the previous section are indeed pure and give us the analytic 
form of the differential equations.

\subsection{The Symbol Alphabet}
\label{sec:AlphabetResults}

Our strategy for determining the symbol alphabet is the same as the one used 
in ref.~\cite{Abreu:2020jxa}, where we used numerical evaluations of the 
differential equations and cut differential equations to reconstruct the 
full symbol alphabet of each topology. The random-direction differential 
equation defined in \cref{eq:randomDirectionDE} will play a central role in 
this procedure.

The first question we can ask about the symbol alphabet is how many letters 
it has.  
To determine the dimension of the symbol alphabet, we compute the matrix 
$C(\epsilon,\vec s)$ at enough random phase-space points $\vec s_k$ so that 
the entries of the matrices $C(\epsilon,\vec s_k)$ become linearly dependent
for different $k$ (we refer the reader to ref.~\cite{Abreu:2020jxa}
for a very detailed description of the approach).
We find
\begin{equation}\label{eq:abcDim}
  \textrm{dim}\left(\mathcal{A}^{[\mzz]}\right)=39\,,\qquad
  \textrm{dim}\left(\mathcal{A}^{[\zmz]}\right)=56\,,\qquad
  \textrm{dim}\left(\mathcal{A}^{[\zzz]}\right)=63\,.
\end{equation}

The next step is to obtain analytic expressions for the letters in the 
alphabet of each topology. We start from the alphabet of the planar 
integrals that was determined in ref.~\cite{Abreu:2020jxa}. This alphabet 
should be completed by the letters obtained by considering all the permutations
of the massless external legs. These permutations are trivial to construct, 
but we must remove the ones that are not independent. 
Through this procedure, we obtain a set of
156 independent letters. We can then verify that the space spanned by the
39 letters of the $\mzz$ alphabet is included in the space spanned by the 
156 letters we constructed by closing the planar alphabet under all
permutations. This is not true, however, for the $\zmz$ and $\zzz$ alphabets,
which means  we are missing some letters for these two topologies.

Up to permutations, there are four missing letters. One of them
appears in a four-point topology and is available in the literature
\cite{Caola:2014lpa, Gehrmann:2015ora}. 
To determine the three missing letters, we analyze the differential equations 
for the new five-point one-mass non-planar integrals depicted in 
\cref{fig_master_int}. We find that the three new letters appear in the
last topology of \cref{fig_master_int}. 
In fact, it is sufficient to study the differential equations for the associated integrals
on their `maximal cut'. In practice, this means
that we can work modulo integrals which do not have all six propagators of this topology,
which greatly simplifies the form of the differential equation.
As a further simplification, we consider the differential equation on
a univariate slice \cite{Abreu:2018zmy,Abreu:2019odu}, that is along
a line in phase-space where the Mandelstam variables depend linearly
on a single variable.
We find that the new letters are simple functions depending 
on the square root of the polynomial $\nsqrt^{(1)}$
defined in \cref{eq:newRoot}. Before listing the remaining
ones, we comment on how we organize the complete symbol alphabet.

As noted in section \ref{sec:masterInt}, 
when computing two-loop five-point one-mass amplitudes
we must consider other hexa-box topologies corresponding to
permutations of the massless external momenta of the 
thee topologies considered here.
In order to obtain the associated letters,
we complete the new letters by including their image under these 
transformations.
We find an alphabet with 204 letters.

Up to permutations, there are three different square roots that appear
in our symbol letters: $\sqrt{\Delta_3}$, $\sqrt{\Delta_5}$ and 
$\sqrt{\nsqrt}$, defined 
respectively in \cref{eq:gram3,eq:gram5,eq:newRoot}. 
While Feynman integrals must be invariant under
a flip of the sign of the square-roots, this invariance might be broken
by the definition of the pure basis. The operations of flipping the sign
of each square root compose to form a group, 
which is known in the mathematics literature as a `Galois group’.
To organize the letters, we choose them to have simple transformation
properties under each element of this group. That is, we choose letters that either
map onto themselves or their reciprocal when the signs of the square roots are flipped.
We find that there are 127 letter which are Galois invariant, and 
77 which transform non-trivially under the Galois group.

As the alphabet is closed under permutations by construction, we can group the
letters into permutation orbits starting from a generating set of letters. We denote 
as $S_4$ the group of permutations of the four massless momenta.
A given generating letter may be invariant under some permutation of a
subset of the massless momenta, and hence it is sufficient to consider equivalence
classes defined by these groups. We denote such a set
of inequivalent permutations as $S_4 / G$, where $G$ is the sub-group of $S_4$
which leaves the given generator invariant. 
$G$ is in general a product of permutation groups, 
which for example we denote as $S_3[3,4,5]$ for the set of permutations of legs 3, 4 and 5.
As in ref.~\cite{Abreu:2020jxa}, we organize the alphabet first by the Galois 
properties of the letters, and then by their mass dimension.
Several letters can be written in a very compact form by using 
the symbol $\mathrm{tr}_\pm$ defined in \cref{eq:trpm}.\footnote{We note a 
subtlety in representing the letters  in terms of
$\mathrm{tr}_{\pm}$. If expanded in terms of Mandelstam invariants, the expression
involves $\mathrm{tr}_5$, and not $\sqrt{\Delta_5}$. These two expressions
behave differently under permutations. In our alphabet (and ancillary files), we
make the replacement $\mathrm{tr}_5 \rightarrow \sqrt{\Delta}_5$ and then build
the remaining letters via permutations.}
With these notational devices in hand, the Galois invariant letters are 
\newcommand{\perms}[2]{\left\{\sigma\left(#1\right) : \sigma \in #2 \right\}}
\newcommand{\permsBig}[2]{\{\sigma\big(#1\big) : \sigma \in #2 \}}
\begin{align}
  \begin{split}
  W_1 &= p_1^2\,, \\[2mm]
  \{ W_2, \ldots, W_5 \} &= \perms{s_{12}}{S_4 / S_3[3, 4, 5]}\,, \\[2mm] 
  \{ W_6, \ldots, W_{11}\} &= \perms{s_{23}}{S_4 / (S_2[2,3] \times S_2[4,5])}\,, \\[2mm]
  \{ W_{12}, \ldots, W_{15}\} &= \perms{2 \, p_1 \cdot p_2 }{S_4 / S_3[3, 4, 5]}\,, \\[2mm] 
  \{ W_{16}, \ldots, W_{27}\} &= \permsBig{2 \, p_2 \cdot (p_3 + p_4)}{S_4 / S_2[3,4]}\,, \\[2mm] 
  \{ W_{28}, \ldots, W_{33}\} &= \permsBig{\mathrm{tr}_+(1\, 2\, 1\, 5)}{S_4 / (S_2[2, 5] \times S_2[3, 4])}\,, \\[2mm] 
  \{ W_{34}, \ldots, W_{45}\} &= \permsBig{\mathrm{tr}_+(1\, 2\, 1\, [4+5])}{S_4 / S_2[4, 5]}\,, \\[2mm] 
  \{ W_{46}, \ldots, W_{57}\} &= \permsBig{\mathrm{tr}_+(1\, [2+3]\, 4\, [2+3])}{S_4 / S_2[2, 3]}\,, \\[2mm] 
  \{ W_{58}, \ldots, W_{69}\} &= \permsBig{\mathrm{tr}_+(1\, 2\, [4+5]\, [2+3])}{S_4 / S_2[4,5]}\,, \\[2mm] 
  \{ W_{70}, \ldots, W_{93}\} &= \permsBig{\mathrm{tr}_+(1\, 2\, 3\, 4) - \text{tr}_+(1\, 2\, 4\, 5)}{S_4}\,, \\[2mm]
  \{ W_{94}, \ldots, W_{117}\} &= \permsBig{\mathrm{tr}_+(1\, 2\, 1\, [1+5]\, 4\,[1+5])}{S_4}\,. 
  \end{split}
\label{eq:GaloisInvariantLetters}
\end{align}
We stress that the letters in \eqref{eq:GaloisInvariantLetters} which make use
of $\mathrm{tr}_+$ are non-trivially invariant under the 
$\sqrt{\Delta_5}\rightarrow -\sqrt{\Delta_5}$ Galois transformation.
The letters with non-trivial Galois properties are
\begin{align}
  \begin{split}
  \{ W_{118}, \ldots, W_{123}\} &=  \perms{\frac{s_{12} + s_{13} + \sqrt{\Delta_3^{(1)}}}{s_{12} + s_{13} - \sqrt{\Delta_3^{(1)}}}}{ S_4 / (S_2[2, 3] \times S_2[4,5]) }\,, \\ 
  \{ W_{124}, \ldots, W_{129}\} &=  \perms{\frac{s_{12} - s_{13} + \sqrt{\Delta_3^{(1)}}}{s_{12} - s_{13} - \sqrt{\Delta_3^{(1)}}}}{ S_4 / (S_2[2, 3] \times S_2[4,5]) }\,, \\ 
  \{ W_{130}, \ldots, W_{137}\} &=  \perms{\frac{\mathrm{tr}_+(1\,2\,3\,4)}{\text{tr}_-(1\,2\,3\,4)}}{\mathcal{S}}\,, \\ 
  \{ W_{138}, \ldots, W_{161}\} &=  \perms{\frac{\mathrm{tr}_+(1\,5\,3\,[1+2])}{\text{tr}_-(1\,5\,3\,[1+2])}}{S_4}\,, \\
  \{ W_{162}, \ldots, W_{185}\} &=  \perms{\frac{s_{12}s_{23} + s_{23} s_{34} - s_{34}s_{45} +s_{45}s_{15} - s_{12}s_{15} + \sqrt{\nsqrt^{(1)}}}{ s_{12}s_{23} + s_{23} s_{34} - s_{34}s_{45} +s_{45}s_{15} - s_{12}s_{15} - \sqrt{\nsqrt^{(1)}}}}{S_4}\,, \\
  \{ W_{186}, \ldots, W_{188}\} &=  \perms{\frac{\Omega^{--} \Omega^{++}}{\Omega^{-+} \Omega^{+-}}}{S_4 / (S_2[2, 3] \times S_2[4, 5] \times S_2[s_{23}, s_{45}])}\,, \\
  \{ W_{189}, \ldots, W_{194}\} &=  \perms{\frac{\tilde{\Omega}^{--} \tilde{\Omega}^{++}}{\tilde{\Omega}^{-+} \tilde{\Omega}^{+-}}}{S_4 / (S_2[3, 4] \times S_2[2,5])}\,, 
  \end{split}
\end{align}
where 
\begin{align}\begin{split}
  \Omega^{\pm \pm} &= s_{12} s_{15} - s_{12} s_{23} - s_{15}s_{45} \pm s_{34} \sqrt{\Delta_3^{(1)}} \pm \sqrt{\Delta_5}\,, \\
  \tilde{\Omega}^{\pm \pm} &= p_1^2 s_{34} \pm \sqrt{\Delta_5} \pm \sqrt{\nsqrt^{(1)}}\,, 
\end{split}\end{align}
and the set of permutations $\mathcal{S}$ is given by
\begin{align}
  \begin{split}
    \mathcal{S} = \big\{&\{1,2,3,4,5\}, \{1,2,3,5,4\}, \{1,2,4,3,5\}, \{1,2,4,5,3\}, \\
                    &\{1,2,5,3,4\}, \{1,3,2,4,5\}, \{1,3,2,5,4\}, \{1,4,2,5,3\} \big\}\,.
  \end{split}
\end{align}
Finally, the square roots in the problem are also letters
\begin{align}
  \begin{split}
  \{ W_{195}, \ldots, W_{197} \} &= \perms{\sqrt{\Delta_3^{(1)}}}{S_4 \ {S_4 / (S_2[2, 3] \times S_2[4, 5] \times S_2[s_{23}, s_{45}])}}\,, \\
  W_{198} &= \sqrt{\Delta_5}\,, \\
  \{ W_{199}, \ldots, W_{204} \} &= \perms{\sqrt{\nsqrt^{(1)}}}{S_4 / (S_2[3, 4] \times S_2[2, 5])}\,. 
  \end{split}
\end{align}

We finish with two comments.
First, the new letters that cannot be obtained from 
the closure of the planar alphabet under permutations
are generated by $W_{58}$, $W_{162}$, $W_{189}$ and $W_{199}$. 
$W_{58}$ appears in four-point integrals \cite{Caola:2014lpa, Gehrmann:2015ora},
and the remaining three letters appear for the first time in the last five-point
topology in \cref{fig_master_int}.
Second, the complete alphabet can be found in the ancillary 
file \texttt{anc/alphabet.m}, 
written explicitly in terms of Mandelstam invariants and the square roots of the polynomials
${\Delta_3}$, ${\Delta_5}$ and $\nsqrt$ and their permutations.
The later are given explicitly in the file \texttt{anc/roots.m}.

\subsection{Analytic Differential Equations from Numerical Samples}
\label{sec:DEFitting}

Once the alphabet has been determined, we take \cref{eq:CAnsatz}
as an ansatz, where we assume the matrices $M_\alpha$ to be matrices
of rational numbers. Using the same numerical evaluation of the differential
equations which were used to determine the dimensions of the symbol alphabet
quoted in \cref{eq:abcDim}, and assuming our ansatz is complete, 
we can determine the $M_\alpha$. This can be done through linear
algebra as detailed in ref.~\cite{Abreu:2020jxa}.
For each of the three non-planar hexa-box topologies, we have
successfully determined
the matrices $M_\alpha$. This confirms that the connections take the form
given in \cref{epsFactorizedDE}, which in turn confirms that the bases 
we have constructed are indeed pure.

The matrices that allow to reconstruct the analytic form of the
differential equations we have obtained can be found
in the ancillary files \texttt{anc/f/f\_connections.m}, 
for $\texttt{f}\in\{\texttt{mzz}, \texttt{zmz}, \texttt{zzz}\}$. 
We provide a  \texttt{Mathematica} example file \texttt{anc/usageExample.m}, 
that assembles the differential equation
for each topology in terms of the alphabet quoted above.

\subsection{Symbols of Non-Planar Hexa-Box Integrals}
\label{sec:strucSymb}

Having confirmed that we have pure bases of master integrals for the three hexa-box topologies, 
and with the analytic differential equations in hand, we can now study some of the analytic
properties of the master integrals by constructing their so-called symbol \cite{Goncharov:2010jf}.

We start by normalizing all master integrals so that their Laurent expansion
around $\epsilon=0$ has no negative powers. It then follows from the form of the canonical differential
equation discussed in \cref{sec:canDiffEq} that
\begin{eqnarray}
\label{eqn:solExpansion}
    {\bf I} = \sum_{i=0} {\bf I}^{(i)} \epsilon^i \,, \qquad 
    {\bf I}^{(i+1)} = \int \sum_{\alpha} M_\alpha  \mathrm{d}\log(W_\alpha) \, {\bf I}^{(i)} \,.
\end{eqnarray}
Furthermore, the derivative of ${\bf I}^{(0)}$ vanishes, which means it has to be a constant vector.
The primitive ${\bf I}^{(n)}$ at arbitrary order $n$, can be written as an iterated integral
\begin{equation}\label{eq:solabs}
{\bf I}^{(n)} = \sum_{\alpha_1, \ldots, \alpha_n} {\bf e}_{\alpha_1, \ldots, \alpha_n} 
\int \mathrm{d}\log W_{\alpha_1} \cdots \mathrm{d} \log W_{\alpha_n}\,,
\end{equation}
and the number of integrations is called the weight of the function, which we note is tied
with the order in the $\epsilon$ expansion.
To obtain the integral functions, we should specify the integration contour and boundary conditions,
which we will discuss in the next section.
Here, we focus on the
integrand of \cref{eq:solabs} which already captures a lot of the analytic structure of the solution.
The symbol associated with these integrals is defined as
\begin{equation}
S[{\bf I}^{(n)}] = \sum_{\alpha_1, \ldots, \alpha_n} 
{\bf e}_{\alpha_1, \ldots, \alpha_n} \left[ W_{\alpha_1}, \cdots, W_{\alpha_n}\right]\,,
\end{equation}
where the coefficients ${\bf e}$ are computed from products of the matrices $M_\alpha$.
If ${\bf I}$ is a vector of master integrals, then its symbol is constrained to satisfy
the first-entry condition~\cite{Gaiotto:2011dt}, which states that the first
entry of all the terms in the symbol tensor (i.e., $W_{\alpha_1}$ in the equation above)
must correspond to a physical channel of the topology. The sets of first entries $\mathcal{F}$ of each of the
three hexa-box topologies in \cref{fig_families_int} are different, and given by
\begin{align}\begin{split}
  \mathcal{F}_{\mzz}=\{\offShellScale,s_{12},s_{23},s_{34},s_{45},s_{15},s_{35},s_{14}\}\,,\\
  \mathcal{F}_{\zmz}=\{\offShellScale,s_{12},s_{23},s_{34},s_{45},s_{15},s_{35},s_{24}\}\,,\\
  \mathcal{F}_{\zzz}=\{\offShellScale,s_{12},s_{23},s_{34},s_{45},s_{15},s_{25},s_{14}\}\,.
\end{split}\end{align}
The first-entry condition for topology $f$ then states that ${\bf e}_{\alpha_1, \ldots, \alpha_n}=0$
if $W_{\alpha_1}\notin \mathcal{F}_{f}$. Equivalently, it states that the weight-zero solution
${\bf I}^{(0)}$, which we recall is a constant, must be in the kernel of all $M_\alpha$ for
$W_{\alpha}\notin \mathcal{F}_{f}$. This turns out to be a surprisingly strong condition
that fully determines ${\bf I}^{(0)}$ for the three hexa-box topologies up to an overall 
normalization (the canonical differential equation is invariant under rescaling of ${\bf I}$
by an $\epsilon$ dependent function).
Imposing the first-entry condition, and using the canonical differential equations we have 
constructed, it is trivial to construct the symbol of all the master integrals in
each of the hexa-box topologies. To illustrate the usage of the differential equations we include
in the ancillary files, we provide a routine to compute the
symbol of the integrals in the example file \texttt{anc/usageExample.m}.

The first-entry condition follows from the fact that the discontinuities of
Feynman integrals should be at physical thresholds~\cite{Gaiotto:2011dt}.
The Steinmann relations impose a further constraint on the analytic structure
of Feynman integrals, as they
state that there should not be double discontinuities on overlapping
channels~\cite{Steinmann, Steinmann2,Cahill:1973qp, Caron-Huot:2016owq, 
Dixon:2016nkn}.  This constrains the first two entries of the symbol of Feynman integrals. 
For each of the hexa-box topologies, the Steinmann relations impose conditions
on different double discontinuities. For $\mzz$ and $\zzz$, 
the forbidden overlapping channels are any pair 
of distinct elements of $\{s_{12},s_{14},s_{15}\}$, corresponding to letters
$\{W_2,W_{4},W_5\}$. 
For $\zmz$, only a single pair of channels is forbidden $\{s_{12},s_{15}\}$,
corresponding to $\{W_2,W_5\}$.
It is easy to verify by computing the symbols at weight 2 that the 
Steinmann relations are satisfied by all master integrals. Nevertheless,
we note a major difference compared to the planar case: for the planar
two-loop five-point integrals considered in ref.~\cite{Abreu:2020jxa},
we observed that the integrals satisfied a stronger version of the
Steinmann relations, called the `extended Steinmann relations' \cite{Caron-Huot:2018dsv}. These
state that the pairs of letters that are forbidden
in the first two entries of the symbol can in fact not appear in 
the $n$-th and $(n+1)$-th entries for any $n$. In ref.~\cite{Abreu:2020jxa},
this stronger version of the Steinmann relations was seen to be a consequence
of the fact that the product of the matrices associated with the
constrained channels vanished, 
and so the letters could never appear next to each other.
For the non-planar hexa-boxes we find that the extended Steinmann relations are
not always satisfied. The situation is as follows:
\begin{itemize}
  \item {\bf $\mzz$}: The extended Steinmann relations are satisfied. 
  Indeed, we find that
  \begin{equation}
    M_2\,M_4=M_4\,M_2=M_2\,M_5=M_5\,M_2=M_4\,M_5=M_5\,M_4=0\,.
  \end{equation}

  \item {\bf $\zmz$}: The extended Steinmann relations are not satisfied.
  By explicit calculation of the symbols through weight 6, we find that
  the master integrals whose symbols involve the sequence 
  $[\ldots,W_2,W_5,\ldots]$
  are at positions $\{1, 2, 3, 8, 9, 10\}$, and the 
  sequence $[ \ldots, W_5,W_2, \ldots]$ appears
  at positions $\{1, 2, 3, 11, 12, 13\}$ of the list of master integrals.

  \item {\bf $\zzz$}: The extended Steinmann relations are satisfied for some
  pairs of channels, but not all. Indeed, we find that
  \begin{equation}
    M_2\,M_5=M_5\,M_2=M_4\,M_5=M_5\,M_4=0\,,
  \end{equation}
  which implies that letters $W_2$ and $W_4$ never appear next to $W_5$. 
  By explicit calculation of the symbols through weight 6, we find that
  the master integrals whose symbols involve the sequence $[\ldots,W_2,W_4,\ldots]$
  are at positions $\{1, 2, 3, 10, 11, 12, 13, 14, 15\}$ 
  and the integrals whose symbols involve the 
  sequence $[\ldots,W_4,W_2,\ldots]$ 
  are at positions $\{1, 2, 3, 4, 5, 6, 7, 8, 9\}$.
\end{itemize}
It would certainly be interesting to further investigate the reasons why
the extended Steinmann relations hold in some cases and not in others, but
we leave this for future work.


\section{Numerical Solution of Differential Equations}
\label{sec:numerics}

\subsection{Summary of the Approach}
\label{sec:diffEqSol}

To solve the differential equations for the non-planar hexa-box topologies 
we will follow the approach of refs.~\cite{Francesco:2019yqt,Hidding:2020ytt}.
In this section we will simply outline the main steps of this approach.
Aside from the two references above, we refer the reader to ref.~\cite{Abreu:2020jxa} 
for a more detailed
discussion in the very closely related context of the solution 
of the planar penta-box topologies.

We consider a vector ${\bf I}(\vec s)$ of pure integrals that
satisfies the differential equation
\begin{equation} \label{eq:diffEqGen}
	 {\rm d} {\bf I}(\vec s)=\eps\,{\bf M}(\vec s)\,{\bf  I}(\vec s)\,.
\end{equation}
We assume the solution is known at the point $\vec s_b$, and our
goal is to compute the solution at a point~$\vec s_e$. To achieve this, 
we consider the one-dimensional path
\begin{equation}\label{eq:path1}
\vec s(t)=  \vec s_b + (\vec s_e-\vec s_b)\,t\,, \qquad  t\in [0,1]\,.
\end{equation}
On this path, the differential equation takes the form
\begin{equation}
\label{eq:DE along contour}
\frac{{\rm d} {\bf I}(t,\epsilon)}{{\rm d}t}=\epsilon\mathbf{A}(t){\bf I}(t,\epsilon)\,,\qquad 
\mathbf{A}(t)=\frac{{\rm d} {\bf M}(\vec s(t))}{{\rm d}t}\,.
\end{equation}

This differential equation can be iteratively solved order by order in 
$\epsilon$.
By normalizing the integrals appropriately, we can ensure that the series around
$\epsilon=0$ has no negative powers, that is
\begin{equation}
    \label{ExactSolutionIteInt t}
    {\bf I}(t,\epsilon) = \sum_{i=0} {\bf I}^{(i)}(t)\,\epsilon^i\,,\qquad
    {\bf I}^{(i)}(t) = \int_0^t \mathbf{A}(t'){\bf I}^{(i-1)}(t')\,{\rm d}t' + \mathbf{c}^{(i)}\,,
\end{equation}
where the integration constants $\mathbf{c}^{(i)}$ are determined by the 
boundary value ${\bf I}(0,\epsilon)$. At each order in~$\epsilon$,
the solution is constructed by patching together locally-valid solutions,
which are themselves written in terms of generalized power series.
The local solution around the point $t_k$ is generically given by
\begin{equation}
	{\bf I}^{(i)}_{k}(t)=\sum_{j_1=0}^\infty 
	\sum_{j_2=0}^{N_{i,k}}\mathbf{c}_{k}^{(i,j_1,j_2)}(t-t_k)^{\frac{j_1}{2}}\log{(t-t_k)}^{j_2}\,,
    \label{eq:SeriesFISingular}
\end{equation}
where $N_{i,k}$ is the maximum power of the logarithms in the local solution
(which is bounded from above by the order $i$ of the $\epsilon$ expansion).
The constants $\mathbf{c}_{k}^{(i,j_1,j_2)}$ are  constrained by continuity conditions between patches and depend on the boundary data
as well as the expansion of $\mathbf{A}(t)$ around the point $t_k$.
To make this approach practical for numerical evaluations, the local
solutions in \cref{eq:SeriesFISingular} are truncated at a finite value of $j_1$.
This value is determined by requiring that the solutions should be valid to a given
numerical accuracy.

As already noted above, we only gave a very brief outline of the approach we use
to solve the differential equations, and we refer the reader to 
refs.~\cite{Francesco:2019yqt,Abreu:2020jxa,Hidding:2020ytt} for more details.

\subsection{Initial Value}
\label{sec:bcDet}

The general solution to the differential equation 
in \cref{eq:diffEqGen} will have branch cuts starting at all the surfaces 
in the space of the Mandelstam variables where ${\bf M}$ is singular. 
Feynman integrals, however, have a more constrained branch-cut structure,
and in particular they should be purely real or purely imaginary in their 
Euclidean region $\mathcal{E}$. The Euclidean region  
associated to each of the three hexa-box topologies is different:
\begin{align}\begin{split}
	\mathcal{E}_{\mzz}&=\{\vec s\in \mathbb{R}^6\,|\,
	\offShellScale<0,s_{12}<0,s_{23}<0,s_{34}<0,s_{45}<0,
	s_{15}<0,s_{35}<0,s_{14}<0\}\,,\\
	\mathcal{E}_{\zmz}&=\{\vec s\in \mathbb{R}^6\,|\,
	\offShellScale<0,s_{12}<0,s_{23}<0,s_{34}<0,s_{45}<0,
	s_{15}<0,s_{35}<0,s_{24}<0\}\,,\\
	\mathcal{E}_{\zzz}&=\{\vec s\in \mathbb{R}^6\,|\,
	\offShellScale<0,s_{12}<0,s_{23}<0,s_{34}<0,s_{45}<0,
	s_{15}<0,s_{25}<0,s_{14}<0\}\,.
\end{split}\end{align}
Requiring that Feynman integrals should be either
purely real or purely imaginary in their associated Euclidean region 
$\mathcal{E}$ implies that there should not be any branch cuts in $\mathcal{E}$,
and this can be used to constrain the initial condition required to solve 
their differential equation. This is
very closely connected to the first-entry condition discussed in 
\cref{sec:strucSymb}, which fully determined the weight 0 solution of the 
differential equation.

To determine the initial value beyond weight 0,
we employ the technique developed in \cite{Abreu:2020jxa}, to which
we refer the reader for further details. Here we simply present a summary 
of the approach.
Let $\vec s_{\mathcal{E},0}$ be the point in $\mathcal{E}$ where
we want to determine the initial condition, and let us
assume that the initial condition is known at order
$\epsilon^{i-1}$. Our goal is then to determine the components of the
vector ${\bf I}^{(i)}(\vec s_{\mathcal{E},0})$. Using the strategy outlined
above, we transport the solution
from $\vec s_{\mathcal{E},0}$ to some other point
$\vec s_{\mathcal{E},1}\in\mathcal{E}$ along a straight line, chosen such that the line
from $\vec s_{\mathcal{E},0}$ to $\vec s_{\mathcal{E},1}$ is fully contained
in $\mathcal{E}$.
The coefficients $\mathbf{c}_{k}^{(i,j_1,j_2)}$ in the local
solutions of \cref{eq:SeriesFISingular} depend linearly on the unknown components of
${\bf I}^{(i)}(\vec s_{\mathcal{E},0})$.
The requirement that there should not be logarithmic branch cuts in
$\mathcal{E}$ means that there should be no logarithms in these local solutions.
That is, the branch-cut constraint amounts to setting to zero any coefficients of the form
$c_k^{(i,j_1,j_2)}$ for which $j_2=1$.
We note that $0\leq j_2\leq 1$ because the initial condition at the previous order
must satisfy the same condition and is assumed to be known,
and we can only generate one power of logarithm per integration.
By collecting all such conditions on the path between
$\vec s_{\mathcal{E},0}$ and $\vec s_{\mathcal{E},1}$, we construct a system of
linear equations for the components of ${\bf I}^{(i)}(\vec s_{\mathcal{E},0})$.
This procedure can then be repeated by considering further points
$\vec s_{\mathcal{E},k}$ and collecting more conditions on the path from 
$\vec s_{\mathcal{E},0}$ to $\vec s_{\mathcal{E},k}$.
In general, for a vector of $n$ Feynman integrals, we can use this procedure to
construct at most $n-1$ independent conditions which determine the vector
${\bf I}^{(i)}(\vec s_{\mathcal{E},0})$ up to an overall normalization. 
We note, however, that since the vector
${\bf I}$ contains many integrals that are already known (for instance, because they correspond
to lower-point topologies), we do not need to collect the maximum number of conditions, but simply
to find the conditions that determine the value of the unknown integrals at $\vec s_{\mathcal{E},0}$.

For the $\zzz$ hexa-box topology, we choose as the initial point
\begin{equation}\label{eq:initZZZ}
	\vec s_{\mathcal{E}_{\zzz},0}= (-13,  -7, -31, -22, -4, -17)\,.
\end{equation}
If we consider the paths from $\vec s_{\mathcal{E}_{\zzz},0}$ to the three points
\begin{align}\begin{split}\label{eq:singPointsZZZ}
	\vec s_{\mathcal{E}_{\zzz},1}=&\left(-\frac{117}{55},-\frac{8}{21},-\frac{68}{139},-\frac{6}{127},
	-\frac{83}{173},-\frac{61}{82}\right)\,,\\
	\vec s_{\mathcal{E}_{\zzz},2}=&\left(-\frac{446}{137},-\frac{31}{119},-\frac{40}{53},-\frac{15}{137},
	-\frac{32}{27},-\frac{149}{96}\right)\,,\\
	\vec s_{\mathcal{E}_{\zzz},3}=&\left(-\frac{104}{61},-\frac{39}{55},-\frac{59}{115},-\frac{21}{184},
	-\frac{1}{2},-\frac{88}{145}\right)\,,
\end{split}\end{align}
we obtain 134 independent conditions. This is the maximal number we could have expected given that
there are 135 master integrals in this topology. To fix the remaining
degree of freedom, we set the master integral corresponding to the
sunrise integral with external mass $q^2$ to be
\begin{equation}
	\textrm{Sr}(q^2)=4(-q^2)^{-2\epsilon}
	\frac{\Gamma(1-\epsilon)^3\Gamma(1+2\epsilon)}{\Gamma(1 - 3 \epsilon)}\,.
\end{equation}
In our conventions, for instance, the $\offShellScale$-sunrise integral
appears in position 135 of the $\zzz$ master integrals.

For the $\mzz$ topology, we choose as the initial point
\begin{equation}\label{eq:initMZZ}
	\vec s_{\mathcal{E}_{\mzz},0}= (-13,-7,-21,-2,-4,-10)\,.
\end{equation}
We then consider the paths from this point to the three points
\begin{align}\begin{split}\label{eq:singPointsMZZ}
	\vec s_{\mathcal{E}_{\mzz},1}=&\left(-\frac{6829}{10},-\frac{14777}{20},-\frac{903}{10},
	-\frac{14677}{20},-\frac{27}{20},-\frac{3389}{5}\right)\,,\\
	\vec s_{\mathcal{E}_{\mzz},2}=&\left(-\frac{4874}{5},-\frac{3913}{4},-\frac{2079}{20},
	-\frac{9407}{10},-\frac{65}{4},-\frac{19426957}{18640}\right)\,,\\
	\vec s_{\mathcal{E}_{\mzz},3}=&\left(-\frac{193817}{20},-\frac{192017}{20},-\frac{147}{2},
	-\frac{191917}{20},-\frac{11}{20},-\frac{38743}{4}\right)\,,
\end{split}\end{align}
and obtain 82 independent conditions. The undetermined initial conditions all correspond to 
known single-scale integrals.

Finally, for the $\zmz$ topology we determined the initial condition at
\begin{equation}\label{eq:initZMZ}
	\vec s_{\mathcal{E}_{\zmz},0}= (-13,-7,-21,-2,-4,-30)\,.
\end{equation}
We consider the paths to
\begin{align}\begin{split}\label{eq:singPointsZMZ}
	\vec s_{\mathcal{E}_{\zmz},1}=&\left(-\frac{155}{128},-\frac{103}{83},-\frac{51}{109},
	-\frac{17}{82},-\frac{69}{197},-\frac{101}{85}\right)\,,\\
	\vec s_{\mathcal{E}_{\zmz},2}=&\left(-\frac{69}{43},-\frac{148}{137},-\frac{12}{77},
	-\frac{57}{89},-\frac{23}{97},-\frac{77}{73}\right)\,,\\
	\vec s_{\mathcal{E}_{\zmz},3}=&\left(-\frac{181}{105},-\frac{79}{88},-\frac{21}{74},
	-\frac{38}{67},-\frac{33}{103},-\frac{89}{93}\right)\,,
\end{split}\end{align}
and collect 62 conditions along the way.
The undetermined integrals are either simple integrals
that can be computed to arbitrary order in $\epsilon$, or
integrals that appear in the $\mzz$ or $\zzz$
hexa-box topologies (sometimes for different permutations of the massless momenta) 
and can thus be computed with the differential equations and
boundary conditions determined for these two topologies.

We note that we have not made an effort to prove whether or not we could have found
other lines in $\mathcal{E}_{\mzz}$ and $\mathcal{E}_{\zmz}$ that would allow
to obtain more conditions for the $\mzz$ and $\zmz$ topologies respectively.
This is certainly an interesting question which we leave for future study.

Having established our strategy to determine the initial values for each integral,
we computed them with two independent implementations.
The first was using the code of ref.~\cite{Hidding:2020ytt}, and the second was
using an in-house implementation that builds upon the same ideas.
With the latter, we obtained the initial conditions with 100 digit precision
and these high-precision evaluations can be found in the ancillary files.
Using the code of ref.~\cite{Hidding:2020ytt}, we were able to validate all
of our initial conditions to at least 25 digits.
We note that, due to the larger number of master integrals and larger alphabet,
the main challenge in this procedure is the determination of the weight-4
boundary conditions for the $\zzz$ topology.

\subsection{Numerical Evaluations in Physical Regions}
\label{sec:physRegionEval}

Having determined the value of the integrals at a point as described in the previous section,
we can then use the approach summarized in \cref{sec:diffEqSol} to obtain
the solutions at arbitrary points in phase-space. With phenomenological applications
in mind, we focus here on the points corresponding to the production of a massive 
vector boson in association with two jets in QCD. The massless partons are assigned
the massless momenta $p_i$, $i=2,\ldots,5$, and the massive momentum $p_1$ is assigned
to the vector boson which we assume to decay, into e.g.~a lepton pair. This requires
that $p_1$ is timelike, that is $p_1^2>0$. There are six different channels of the form
\begin{equation}
p_i+p_j \rightarrow p_1+p_k+p_l\,,
\end{equation}
where $i,j,k,l$ take distinct values in $\{2,3,4,5\}$. In \cref{tab:regions} we present
the signs of the Mandelstam variables in $\vec s$ for each of the channels.

\begin{table}[]
\centering
 \begin{tabular}{| c | c | c | c | } 
\hline
	Initial State  	& $	>0	$ & $<0	$\\\hline
	$2,3$ & $ s_{23},s_{45},s_{15},\offShellScale $&$ s_{12},s_{34} $ \\   
	$2,4$		& $ s_{15},\offShellScale 		$&$ s_{12},s_{23},s_{34},s_{45}   $ \\
	$2,5$		& $ s_{34},\offShellScale 		$&$ s_{12},s_{23},s_{45},s_{15}   $ \\ 
	$3,4$		& $ s_{12},s_{34},s_{15},\offShellScale $&$ s_{23},s_{45} 		  $  \\
	$3,5$		& $ s_{12},\offShellScale 		$&$ s_{23},s_{34},s_{45},s_{15}	  $	\\
	$4,5$		& $ s_{12},s_{23},s_{45},\offShellScale $&$ s_{34},s_{15}         $ \\\hline
  \end{tabular}
 \caption{Signs of the Mandelstam variables in $\vec s$ for the production of a massive vector 
 boson (of momentum $p_1$) in association with two jets in QCD. We consider the channels
 corresponding to any pair of massless momenta in the initial state.}
 \label{tab:regions}
\end{table}

To demonstrate that we are indeed able to evaluate the master integrals in
phase-space regions of physical interest, we choose a point in each of the physical regions
defined in \cref{tab:regions},
\begin{align}\begin{split}
\label{eq:p_phys}
\vec{s}_{\text{ph-}1} &= \left(  \frac{137}{50},   - \frac{22}{5},    
\frac{241}{25},   - \frac{377}{100},    \frac{13}{50},    \frac{249}{50} \right)\,,\\
\vec{s}_{\text{ph-}2} &=  \left(  \frac{137}{50},   - \frac{22}{5},   
- \frac{91}{100},   - \frac{377}{100} ,  - \frac{9}{10} ,   \frac{249}{50} \right)\,,\\
\vec{s}_{\text{ph-}3} &=  \left(  \frac{137}{50},   - \frac{22}{5},  
- \frac{91}{100} ,   \frac{13}{50},   - \frac{9}{10} ,  - \frac{9}{4}\right)\,, \\
\vec{s}_{\text{ph-}4} &=  \left(  \frac{137}{50},    \frac{357}{50},  
 - \frac{91}{100},    \frac{241}{25} ,  - \frac{9}{10} ,   \frac{249}{50}\right)\,, \\
 \vec{s}_{\text{ph-}5} &= \left(  \frac{137}{50},    \frac{357}{50},  
  - \frac{91}{100} ,  - \frac{161}{100} ,  - \frac{9}{10} ,  - \frac{9}{4}\right)\,, \\
 \vec{s}_{\text{ph-}6} &= \left(  \frac{137}{50},    \frac{357}{50},  
   \frac{13}{50},   - \frac{161}{100} ,   \frac{241}{25},   - \frac{9}{4}\right)\,. 
\end{split}\end{align}
These points are the same we have used in 
ref.~\cite{Abreu:2020jxa} for the evaluation of the planar topologies
and have been chosen at random in each of the regions. We include high-precision
evaluations  at these points in our ancillary files, which were obtained
with the code of ref.~\cite{Hidding:2020ytt}. The numbers we provide
are correct to 100 digits, and can be directly used as boundary conditions for evaluations
in each of the physical regions. We close this section by noting that some points
cannot be reached with a straight line from the Euclidean boundary point as it would
require to analytically continue through a non-physical threshold. Instead, we take
an indirect path built from two line segments that avoids this problem.

\subsection{Validation}

Let us first describe the validation of the Euclidean initial values discussed
in~\cref{sec:bcDet}. 
First, as already noted, the results we present were computed with an in-house
implementation of the algorithm of ref.~\cite{Francesco:2019yqt}, and validated
at lower precision with their evaluation with the code
of~ref.~\cite{Hidding:2020ytt}.
Second, we have also computed them with an independent in-house implementation
of the algorithm of ref.~\cite{Francesco:2019yqt}.
Third,
the results for the $\mzz$ topology and selected integrals in the $\zmz$ and $\zzz$  topologies were 
reproduced by the authors of ref.~\cite{Papadopoulos:2019iam}.
Finally, we have used the code we have developed for the determination of the high-precision
initial values to extend the analysis to weight five and verify that our weight-four numerical
results guarantee the absence of non-physical branch cuts inside the Euclidean region.

The high-precision evaluations at the physical points of ref.~\cref{eq:p_phys} have been
obtained with the publicly available code of ref.~\cite{Hidding:2020ytt}, and can thus be easily reproduced.
As a consistency check we verified that we
obtain the same value at each point independently of which point is used as 
initial value.
Finally, we have performed lower-precision comparisons with the second in-house
implementation of the algorithm of~ref.~\cite{Francesco:2019yqt}.


\section{Conclusions}
\label{sec:Conclusions}

In this paper we have taken an important step towards completing the calculation
of the full set of two-loop master integrals with one massive and four massless legs.
The calculation of the three distinct hexa-box non-planar topologies was performed
with well established and tested techniques: after constructing
a pure basis of master integrals, we obtained their differential equation and 
solved them in terms of generalized power series.

While the techniques we use are well established, the complexity of the calculation
is noteworthy. Indeed, in the most complicated topology the differential equation
contains 135 integrals and the dimension of the symbol alphabet is 63.
To handle this complexity when constructing the differential equations, we find
it important to leverage the power of an approach based on
ans\"atze and numerical samples, and to consider differential equations on
maximal cuts.
The results we obtain for the pure basis and the symbol alphabet
are particularly compact. The alphabet itself is generated by appropriate
permutations of only 21 elements, with 204 letters in total.
An interesting feature compared to the planar alphabet, is the appearance of a new
type of square-root which leads to new letters with non-trivial Galois-group
properties.

Having obtained the differential equations for each of the three non-planar hexa-box
topologies, we computed their symbols. As expected, the Steinmann relations
are satisfied, but we find that the extended Steinmann relations are in general
not satisfied in the $\zmz$ and $\zzz$ topologies. It would certainly be
interesting to understand why this is the case, as this would also shed new
light on why these relations hold for planar integrals.

In order to obtain numerical values for the master integrals from their 
differential equations, we must determine them at a point and use this evaluation
as a boundary condition for the differential equation. 
Once again, the large number of master integrals and symbol letters makes this
a non-trivial problem. We developed our own dedicated code to compute a high-precision initial
value for all the master integrals in their respective Euclidean region.
This is achieved
by imposing that there are no non-physical singularities in this region
of phase-space. Imposing this condition allowed us to obtain numerical evaluations
valid to more than 100 digits, which can then be used to obtain numerical 
values for the master integrals in all regions of phase-space. As an example, 
we also provide high-precision evaluations in the six different physical regions
corresponding to the production of a massive particle in association with 
two jets in QCD.

Our results are a new essential ingredient for the calculation of the
two-loop corrections for very important processes, such as the production
of a Higgs boson or a massive vector boson in association with two jets at
hadron colliders.
The analytic
insight we gained from studying the symbols of the non-planar hexa-box master integrals
will be crucial in computing the two-loop amplitudes for these processes,
and the ability to numerically evaluate the integrals in all regions of phase
space makes our results usable for computing theoretical predictions
to physical observables. 
Finally, the results presented in this paper will be very important in evaluating the
remaining topologies required to complete the calculation of the 
full set of two-loop master integrals with one massive and four massless legs.

\section*{Acknowledgments}
We wish to thank Martijn Hidding, Costas Papadopoulos, Michael Ruf and Nikolaos
Syrrakos for discussions.
The work of B.P.~has been supported by the French Agence Nationale
pour la Recherche, under grant ANR--17--CE31--0001--01.
This project has received funding from the European’s Union Horizon 2020
Research and Innovation Programme under grant agreement number 896690, project
‘LoopAnsatz’.
W.T.'s work is funded by the German Research Foundation (DFG) within the Research Training Group GRK 2044.
The authors acknowledge support by the state of
Baden-W\"urttemberg through bwHPC.

\bibliography{main.bib}
\end{document}